\documentclass[10pt,a4paper,twocolumn,showpacs,amsmath,pra,amssymb,nofootinbib]{revtex4} 
\usepackage[T1]{fontenc}
\usepackage[latin9]{inputenc} 
\usepackage{amssymb,amsmath}
\usepackage{amsbsy}
\usepackage[amssymb]{SIunits}
\usepackage{graphicx}
\usepackage{float}
\usepackage{mathrsfs}
\usepackage{times}
\usepackage{ifthen}
\usepackage{xspace}

\newcommand{\diff}{\mathrm{d}}
\newcommand{\intinf}{\int_{-{\infty}}^{\infty}}
\newcommand{\intzinf}{\int_{0}^{\infty}}

\newcommand{\av}[1]{\ensuremath{\left\langle #1 \right\rangle}\xspace}

\newcommand{\pdiffal}[3]{\left[{\partial #1}/{\partial #2}\right]_{#3}}

\newcommand{\tdiffl}[2]{{\diff #1}/{\diff #2}}

\newcommand{\nbe}{\ensuremath{\bar{n}_{\mathrm{BE}}}\xspace}

\newcommand{\ii}{\ensuremath{\mathrm{i}}\xspace}
\newcommand{\com}[1]{\ensuremath{\left[#1\right]}\xspace}
\newcommand{\norm}[1]{\ensuremath{\left|#1\right|}\xspace}
\newcommand{\myatop}[2]{\genfrac{}{}{0pt}{}{#1}{#2}\xspace}

\newcommand{\fracl}[2]{\ensuremath{{#1}/{#2}}\xspace}

\newcommand{\fracb}[2]{\left(\frac{#1}{#2}\right)}

\newcommand{\half}[1][1]{\frac{#1}{2}\xspace}

\newcommand{\ce}{\ensuremath{\mathrm{ce}}\xspace}

\newcommand{\ah}{\ensuremath{{\hat{a}}^{\phantom{\dagger}}}\xspace}

\newcommand{\ahd}{\ensuremath{{\hat{a}}^\dagger}\xspace}

\newcommand{\deh}{\ensuremath{\hat{\delta}}\xspace}
\newcommand{\dehs}{\ensuremath{{\hat{\delta}}^2}\xspace}
\newcommand{\dehd}{\ensuremath{{\hat{\delta}}^\dagger}\xspace}
\newcommand{\dehds}{\ensuremath{{\hat{\delta}}^{\dagger 2}}\xspace}

\newcommand{\nh}{\ensuremath{{\hat{n}}}\xspace}
\newcommand{\Kh}{\ensuremath{\hat{K}}\xspace}
\newcommand{\Kq}{\ensuremath{\hat{K}_\mathrm{Q}}\xspace}
\newcommand{\Khd}{\ensuremath{\hat{K}^\dagger}\xspace}
\newcommand{\Hh}{\ensuremath{\hat{H}}\xspace}
\newcommand{\Hl}{\ensuremath{\Hh_\mathrm{latt}}\xspace}

\newcommand{\Lh}{\ensuremath{\hat{\mathcal{L}}}\xspace}
\newcommand{\Nh}{\ensuremath{\hat{N}}\xspace}

\newcommand{\Sh}[1]{\ensuremath{\hat{S}_{#1}}\xspace}

\newcommand{\Psih}[1][{}]{\ensuremath{\hat{\Psi}}\xspace}
\newcommand{\Psihd}[1][{}]{\ensuremath{\hat{\Psi}^\dagger}\xspace}
\newcommand{\Psihdd}[1][{}]{\ensuremath{\hat{\Psi}^{\dagger'}}\xspace}
\newcommand{\Psihr}[1][{}]{\ensuremath{\hat{\Psi}(\br_{#1})}\xspace}
\newcommand{\Psihdr}[1][{}]{\ensuremath{\hat{\Psi}^\dagger(\br_{#1})}\xspace}

\newcommand{\psit}{\ensuremath{{\tilde{\psi}^{\vphantom{\dagger}}}}\xspace}
\newcommand{\psitd}{\ensuremath{{\tilde{\psi}^\dagger}}\xspace}
\newcommand{\psitr}{\ensuremath{\psit(\br)}\xspace}
\newcommand{\psitdr}{\ensuremath{\psitd(\br)}\xspace}
\newcommand{\alh}{\ensuremath{{\hat{\alpha}^{\vphantom{\dagger}}}}\xspace}
\newcommand{\alhd}{\ensuremath{{\hat{\alpha}^{{\dagger}}}}\xspace}
\newcommand{\pst}[1]{{#1}^{\vphantom{\dagger}}\xspace}
\newcommand{\st}[1]{{#1}^{\vphantom{\dagger}*}\xspace}

\newcommand{\br}{\ensuremath{\mathbf{r}}\xspace}

\newcommand{\bR}{\ensuremath{\mathbf{R}}\xspace}

\newcommand{\bp}{\ensuremath{\mathbf{p}}\xspace}
\newcommand{\bk}{\ensuremath{\mathbf{k}}\xspace}

\newcommand{\bzero}{\ensuremath{\mathbf{0}}\xspace}

\newcommand{\rb}{\ensuremath{\bar{r}}\xspace}

\newcommand{\xb}{\ensuremath{\bar{x}}\xspace}
\newcommand{\yb}{\ensuremath{\bar{y}}\xspace}
\newcommand{\zb}{\ensuremath{\bar{z}}\xspace}

\newcommand{\ob}{\ensuremath{\bar{\omega}}\xspace}

\newcommand{\dbr}{\diff\br}

\newcommand{\dbk}{\diff\bk}

\newcommand{\nc}{\ensuremath{n_c}\xspace}
\newcommand{\nt}{\ensuremath{\tilde{n}}\xspace}
\newcommand{\ntd}[2][{}]{\ensuremath{\tilde{n}_{#2}^{#1}\xspace}}
\newcommand{\ncd}[1][{}]{\ensuremath{n_c^{#1}\xspace}}

\newcommand{\Nc}{\ensuremath{N_c}\xspace}
\newcommand{\Nt}{\ensuremath{\tilde{N}}\xspace}

\newcommand{\vl}[1][{\br}]{\ensuremath{V_\mathrm{latt}(#1)}\xspace}
\newcommand{\vtn}{\ensuremath{V_\mathrm{tr}}\xspace}
\newcommand{\vt}[1][{\br}]{\ensuremath{\vtn(#1)}\xspace}

\newcommand{\gt}{\ensuremath{g_\mathrm{tr}}\xspace}
\newcommand{\ga}{\ensuremath{g_\mathrm{LDA}}\xspace}

\newcommand{\kb}{\ensuremath{k_B}\xspace}
\newcommand{\kt}{\ensuremath{\kb T}\xspace}

\newcommand{\Rb}[1][{}]{\ensuremath{\mathrm{^{87}_{#1}Rb}}\xspace}
\newcommand{\Na}[1][{}]{\ensuremath{\mathrm{^{23}_{#1}Na}}\xspace}

\newcommand{\hwb}{\ensuremath{\hbar\ob}\xspace}

\newcommand{\wR}{\ensuremath{\omega_R}\xspace}

\newcommand{\dc}{\ensuremath{{}^{'}}\xspace}

\newcommand{\ns}[1][{}]{\ensuremath{N_{s\ifthenelse{\equal{#1}{}}{}{,#1}}}\xspace}  

\newcommand{\mufs}{{\mu_\mathrm{fs}}\xspace}

\newcommand{\Lnh}{\ensuremath{\mathcal{L}}\xspace}
\newcommand{\sj}{\sum_{j}}
\newcommand{\csp}{,\hspace{0.5cm}}

\newcommand{\erj}{E_{R,j}}

\newcommand{\Kbm}[1][{b}]{\ensuremath{K_{#1}^{\mathrm{min}}}\xspace}

\newcommand{\Ele}{\ensuremath{E_{\mathrm{cross}}}\xspace}
\newcommand{\gp}{Gross-Pitaevskii\xspace}

\newcommand{\bsplit}{\right.\notag\\&\left.}

\newcommand{\bz}{\ensuremath{\mathrm{BZ}}\xspace} 
\newcommand{\vi}{\ensuremath{v_i}\xspace} 

\newcommand{\unif}{translation\-ally-invariant\xspace}
\newcommand{\Unif}{Translationally-invariant\xspace}
\renewcommand{\kb}{\ensuremath{k}\xspace}
\renewcommand{\csp}{,\hspace{0.05cm}}
\renewcommand{\Ele}{\ensuremath{E_{\mathrm{cr}}}\xspace}

\newcommand{\afn}[2][{}]{#1\footnote{#2}\xspace} 
\begin{document}
\title{Finite temperature theory of superfluid bosons in optical lattices}
\author{D. Baillie and P. B. Blakie}
\affiliation{Department of Physics, Jack Dodd Centre for Quantum Technology, University of Otago, P.O. Box 56, Dunedin, 9016 New Zealand}
\pacs{67.85.Hj, 03.75.Hh, 05.30.Jp }
\date{\today}
\begin{abstract}
A practical finite temperature theory is developed for the superfluid regime of a weakly interacting Bose gas in an optical lattice with additional harmonic confinement. We derive an extended Bose-Hubbard model that is valid for shallow lattices and when excited bands are occupied. Using the Hartree-Fock-Bogoliubov-Popov mean-field approach, and applying local density and coarse-grained envelope approximations, we arrive at a theory that can be numerically implemented accurately and efficiently.  We present results for a three-dimensional system, characterizing the importance of the features of the extended Bose-Hubbard model and compare against other theoretical results and show an improved agreement with experimental data.
\end{abstract}
\maketitle
\section{Introduction}
Bosonic atoms confined in an optical lattice are a remarkably flexible system for exploring many-body physics  \cite{Anderson1998a,Burger2001a,Greiner02b,Hensinger01,Morsch2003a,Orzel2001a,Spielman2006a,Morsch06,Lewenstein07,Bloch08b,Yukalov09}, in which strongly correlated physics can be explored, for example, through the superfluid to Mott-insulator transition \cite{Greiner02a}.
In the superfluid regime, a Bose-Einstein condensate exists and experiments
have explored its properties, such as coherence \cite{Orzel2001a,Greiner2001a,Morsch2002a,Spielman2006a},
collective modes \cite{Fort2003}, and transport \cite{Fertig2005,Burger2001a,Fallani2004b}.
To date few experiments have considered the interplay between the condensate and thermal components that occurs at finite temperatures \cite{Fort2003,Greiner2001a}. Indeed, quantitative experimental studies of the finite temperature regime have been hampered by the lack of an accurate method for performing thermometry in the lattice. Recent experimental work has overcome this issue  \cite{McKay09} (also see \cite{Trotzky2009a}) and finite temperature properties will undoubtedly receive increased interest in the near future. 

A unique feature of many-body physics with ultra-cold atoms is the opportunity to start from a complete microscopic theory and perform \textit{ab initio} calculations that can be directly compared with experiments.
 In the deep lattice and low temperature limits, bosonic atoms in an optical lattice provide a precise realization of the Bose-Hubbard model \cite{Jaksch98}, originally proposed as a toy model for condensed matter physics \cite{Fisher89}. However, there is a wide regime of experimental interest in which the approximations central to the Bose-Hubbard model (nearest neighbor tunneling, local interactions, and neglect of excited bands) are not valid. In such regimes it is necessary to go beyond the Bose-Hubbard model to furnish an accurate description of the physical system.
  
Theoretical understanding of the properties of bosons in optical lattices is still emerging, and accurate modeling is made difficult by the combined harmonic lattice potential used in experiments, which leads to a complex spectrum, even in the absence of interactions \cite{Hooley04,Viverit04,Rey05,Blakie07b}.
One approach is to use quantum Monte Carlo calculations  which, in principle, fully include thermal fluctuations and quantum correlations. Applications of this approach have mainly been to the Bose-Hubbard model
\cite{Kashurnikov2002a,Wessel04,Kato2009a}, although recently a continuous space algorithm has also been developed for the full lattice potential \cite{Sakhel2009a}.
Mean-field methods provide an approximate treatment that is  much simpler to use and are applicable in the superfluid regime where only weak correlations arise from inter-particle interactions.
Extensive studies of the harmonically trapped gas have demonstrated that the Hartree-Fock-Bogoliubov-Popov (HFBP) mean-field theory \cite{Griffin96} provides a capable description of thermodynamic properties \cite{Dalfovo99}, that agrees well with experiments \cite{Gerbier04,Gerbier2004b}.
The development of similar mean-field theories for the lattice system has been much more limited:  HFBP calculations have been performed for one-dimensional lattice systems in the continuous \cite{Arahata2008a,Arahata2009a} and Bose-Hubbard limit \cite{Rey03,Wild06}, and Duan and coworkers have developed a local density version for the three-dimensional Bose-Hubbard model in \cite{Yi07,Lin08}. 
 To obtain a theory suitable for direct experimental comparison over a broad parameter regime, it is necessary to go beyond the approach in Refs.~\cite{Yi07,Lin08} to obtain a formalism valid for shallow lattices and when excited bands are occupied.

In this paper we develop a HFBP formalism, based on an extended Bose-Hubbard model that includes beyond nearest neighbor tunneling, excited band occupation, interactions between bands and we discuss an approximate treatment of off-site interactions. 
An important aim of our work is to provide a formalism suitable for efficient numerical implementation.  To achieve this we make use of a local density approximation (LDA), that accounts for beyond nearest neighbor tunneling and excited bands, and we develop an envelope approximation that simplifies the treatment of a general anisotropic harmonic confinement to a problem with one independent spatial dimension. Combined, the LDA and envelope approximations allow us to realize an efficient and practical numerical formulation. We show under what conditions it reduces to the simplified theory in Refs.~\cite{Yi07,Lin08} and we numerically investigate the features of our formalism.
 
In section \ref{c:bosonol} we derive the many-body Hamiltonian for bosons in an optical lattice with two body interaction, which we convert to the extended Bose-Hubbard Hamiltonian. We make HFBP mean-field approximations to this in section \ref{c:mfield}. We diagonalize the mean-field Hamiltonian in the LDA, and compare our implementation to that of \cite{Yi07,Lin08} in section \ref{c:lda}. We derive results on the rich structure of the LDA combined harmonic lattice density of states in section \ref{c:dos}, which we compare to the full diagonalization of the non-interacting Hamiltonian. 
In section \ref{c:numericalimpl} we show some important features of our numerical implementation and present numerical results from our model in section \ref{c:results}. We compare our predictions of thermal properties with results from the full diagonalization for the ideal gas and with limited experimental results available. We consider the significance of beyond nearest-neighbor hopping and excited bands and illustrate the properties of our model. In the appendices, we consider the extended Bose-Hubbard parameters, including an approximate interpolative scheme for off-site interactions.  \enlargethispage{1cm}\vspace{-0.4cm}
\section{Bosons in optical lattices\label{c:bosonol}}
\subsection{Lattice potential and units\label{s:lattpot}}
We consider an optical lattice formed by orthogonal standing waves, created by two opposing lasers in each direction. The laser wavelength $\lambda_j$ (in direction $j$) is off-resonant with respect to an atomic transition. The resulting potential in $d$\label{p:d} dimensions, up to an additive constant, is:
\begin{align}
  \vl \equiv \sum_{j=1}^d V_j\sin^2\left(\frac{\pi r_j}{a_j}\right), \label{e:vlatt}
\end{align}
where $V_j$ is the lattice depth and $a_j\equiv\lambda_j/2$ is the lattice spacing in direction $j$. Most of our results can be generalized to the non-separable lattice by adjusting the density of states we introduce in section \ref{c:dos}. We avoid doing this for notational simplicity. 

Except where specifically stated otherwise, our results are generally valid for non-cubic lattices and lower-dimensional systems\afn[.]{However, we do not consider quasi-reduced-dimensional systems, where some directions are partially accessible, i.e. $\kt$ is of the order of the level spacing.} By a cubic lattice, we mean the underlying Bravais lattice has cubic symmetry (or the equivalent in lower dimensions, such as the square case) and that the lattice spacings, $a_j$, and depths, $V_j$, are the same in each axial direction. 
This is the regime of most 3D experiments 
\cite{Greiner02a,Greiner02b,Folling05,Gerbier05a,Gerbier05b,Gerbier06,Schori04,Xu05,Xu06}.

We will generally present results in recoil units, with the unit of length being $a_j/\pi$ and the unit of energy $E_R \equiv h^2/8ma^2$ where $m$ is the atomic mass and $a\equiv \prod_j a_j^{1/d}$.\vspace{-0.4cm}
\subsection{Harmonic-trap potential\label{s:trappot}}
Experimentally, atoms are subject to a crossed optical dipole \cite{Xu05, Xu06} potential (due to the focused lasers used to make the lattices) and often a magnetic trap also \cite{Folling05,Greiner02b}. These effects are well described by introducing an additional potential that is approximately harmonic in form, i.e.
\begin{align}
  \vt \equiv \half m\sum_{j=1}^d  \omega_j^2 r_j^2,
\end{align}
where $\omega_j$ is the harmonic trap frequency in direction $j$. In 3D experiments, the trap is often spherical or cylindrically symmetric (e.g. $\omega_x=\omega_y\ne \omega_z$). We consider the general anisotropic case in $d$ dimensions. We consider both the lattice with $\vt=0$, which we call the `\unif lattice', and the experimentally relevant combined harmonic trap and optical lattice potential, which we call the `combined harmonic lattice'.

In typical experiments \cite{Greiner02a,Greiner02b,Schori04,Xu05,Xu06}, we find the harmonic trapping frequencies to be generally between $2\pi \times 18\,\hertz$ and $2\pi \times 155\,\hertz$, giving $\omega/\wR$ between $0.005$ and $0.02$ where $\omega \equiv \prod_j\omega_j^{1/d}$ and $\wR \equiv \fracl{E_R}{\hbar}$ is the recoil frequency.
\subsection{Many-body Hamiltonian}
In this work we consider only bosons, with field operator \Psihr\label{p:Psihr} such that \cite{Fetter71}:
\begin{align}
  \com{\Psihr,\Psih(\br')} = 0\csp\com{\Psihr,\Psihd(\br')}= \delta(\br-\br'). \label{e:psicom}
\end{align}
In the ultra-cold regime, a dilute gas of bosons is described by the Hamiltonian \cite{Huang57}:
\begin{align}
  \Hh &= \int\dbr\Psihd \left[\Hl + \vt\right]\Psih 
  + \frac{g}{2} \int\dbr \Psihd \Psihd\Psih\Psih, \label{e:contactH}
\end{align}
where $\Hl \equiv -\fracl{\hbar^2\nabla^2}{2m} + \vl$, $g \equiv 4\pi \hbar^2a_s/m$ and $a_s$ is the s-wave scattering length.
\subsection{Wannier basis}
We expand the boson field operators in a basis of the Wannier functions of the non-interacting \unif lattice, $w_b(\br-\bR_i)$, where $b$ is the band index and $\bR_i$ is the lattice site position (see appendix \ref{a:appwannier}), so that we have (as in \cite{Scarola05}):
\begin{align}
  \Psihr = \sum_{b,i} \ah_{b,i} w_b(\br-\bR_i)
  ,\label{e:Psih}
\end{align}
where $\ah_{b,i}$ is the bosonic destruction operator for an atom in band $b$ at site $i$. We note that $b$ and $i$ are discrete $d$-dimensional vectors. For convenience, we shall refer to the ground band as $b=0$. The Wannier basis is a localized basis for sufficiently deep lattices but, for a given lattice depth, there is less localization for excited bands (see appendix \ref{a:appwannier}).
Using a localized basis significantly simplifies the treatment of interactions when off-site interactions are ignored.

The Wannier states are `quasi-stationary', since they are not eigenstates of \Hl, so that there are transitions between the different Wannier  states in the same band due to the single-particle evolution. In particular, the matrix element for hopping from site $\bR_{i'}$ to site $\bR_{i}$ for band $b$ is defined as:
\begin{align}
     J_{b,i,i'} &\equiv  - \int\dbr\, w^*_{b}(\br-\bR_i)\Hl w_{b}(\br-\bR_{i'}). \label{e:jdef}
\end{align}
There is no inter-band hopping (see \eqref{e:Jn1n2}) with the (non-interacting, \unif lattice) definition of the Wannier functions we are using. A change of variables in \eqref{e:jdef} shows that this formula is dependent on $\bR_i$ and $\bR_{i'}$ only through the difference $\bR_i - \bR_{i'}$. Considering the importance of beyond nearest neighbor hopping, we note that the ground-band next-nearest-neighbor hopping matrix element is as much as $25\%$ of its nearest-neighbor counterpart at $V_j=0$, but decreases rapidly with increasing $V_j$, and that beyond next-nearest-neighbor hopping is less significant, as shown in Fig.~\ref{f:jdj} in appendix \ref{a:appwanhop}.
\subsection{Extended Bose-Hubbard Hamiltonian\label{s:exbh}}
We now express the Hamiltonian in terms of the operators $\ah_{b,i}$ by inserting \eqref{e:Psih} into \eqref{e:contactH} and we consider the resulting terms in this section.

We assume the trap is slowly varying relative to the lattice spacings $a_j$ so that:
\begin{align}
          &\int\dbr\, \vt w^*_{b}(\br-\bR_i) w_{b'}(\br-\bR_{i'}) \notag\\
        \approx \vi &\int\dbr\, w^*_{b}(\br-\bR_i) w_{b'}(\br-\bR_{i'}) 
        = \vi  \delta_{bb'}\delta_{ii'},\label{e:videf}
\end{align}
where $\vi \equiv \vt[\bR_i]$\label{p:vi}. In this work, we will always use the local energy form \eqref{e:videf} to represent the harmonic trap. However, there are approximations involved in \eqref{e:videf} which we consider in appendix \ref{a:trap}. We define the total number operator:
\begin{align}
   \Nh &\equiv \int\dbr\,\Psihdr\Psihr 
   = \sum_{b,i}\nh_{b,i},
\end{align}
where $\nh_{b, i} \equiv \ahd_{b,i}\ah_{b,i}$\label{p:nhbi}. Then, expressing the Hamiltonian in the grand-canonical distribution to conserve total particle number, $\Kh \equiv \Hh - \mu \Nh$:
\begin{align}
  \Kh &= \sum_{b,i}\left[-\sum_{i'}\left(J_{b,i,i'}\ahd_{b,i}\ah_{b,i'} \right) + \nh_{b,i}(\vi-\mu)\right] \notag\\
  &+ \frac{1}{2}\sum_{\myatop{i_1,i_2,i_3,i_4}{b_1,b_2,b_3,b_4}}\hspace{-0.2cm}\ahd_{b_1,i_1}\ahd_{b_2,i_2}\ah_{b_3,i_3}\ah_{b_4,i_4}U_{\myatop{i_1,i_2,i_3,i_4}{b_1,b_2,b_3,b_4}} ,\label{e:Kh}
\end{align}
where $
%
  U_{\myatop{i_1,i_2,i_3,i_4}{b_1,b_2,b_3,b_4} } \equiv g\int\dbr\, w^*_{b_1}(\br-\bR_{i_1})w^*_{b_2}(\br-\bR_{i_2})w_{b_3}(\br-\bR_{i_3})w_{b_4}(\br-\bR_{i_4})$. 
%
 If we restrict to on-site interactions (justified in a deep lattice by the Wannier state locality),  \eqref{e:Kh} reduces to $\Kh = \sum_i \Kh_i$ where:
\begin{align}
  \Kh_i &\equiv \sum_{b}\left[-\sum_{i'}\left(J_{b,i,i'}\ahd_{b,i}\ah_{b,i'} \right) + \nh_{b,i}(\vi-\mu)\right] \notag\\
  &+ \frac{1}{2}\sum_{{b_1,b_2,b_3,b_4}} \ahd_{b_1,i}\ahd_{b_2,i}\ah_{b_3,i}\ah_{b_4,i}U_{\myatop{i,i,i,i}{b_1,b_2,b_3,b_4}}, \label{e:Khi}
\end{align}
\enlargethispage{1cm} (this interaction term has previously been stated by \cite{Isacsson05}). We retain a smaller set of interaction parameters, i.e.:
\begin{align}
   U_{bb'} \equiv g\int\dbr\, \norm{w_{b}(\br)w_{b'}(\br)}^2. \label{e:Ubb} 
\end{align}
which is a good approximation in the typical experimental regime, where the interaction parameters are small compared to the band-gap energy scale so that we may ignore collisional couplings between bands in the many-body state. This approximation would need to be revised in the vicinity of a Feshbach resonance (e.g. see \cite{Diener06}), but this is beyond our scope here.

We derive an approximation scheme for off-site interactions in appendix \ref{a:intcoeff}. The result is a modification of the interaction coefficients. As discussed in appendix \ref{a:intcoeff}, if we use the all-site interaction coefficients in our model at $V_j=0$, with appropriate interpretation of the number densities, our model is exactly the same as existing no-lattice models. For the non-condensate, we find that the effects of off-site interactions are negligible for $V_j\gtrsim 5E_R$. Formulating a consistent theoretical description in the shallow lattice limit is fraught for a Wannier state approach, because these states are delocalized in this regime; some work in the shallow lattice has been reported \cite{Zobay06}. However, our off-site interaction coefficients provide a useful interpolation scheme which is accurate in the no-lattice case and for moderate to deep lattices. For the condensate, interference between sites, mediated by the tails of distant Wannier states, can reduce the interaction coefficient, as discussed in appendix \ref{a:intcoeff}. All of our work other than appendix \ref{a:intcoeff} uses on-site interaction coefficients.

Other extended Bose-Hubbard work has used various simplifications of \eqref{e:Kh}: the use of nearest-neighbor hopping and nearest-neighbor interactions \cite{Scarola05}; the use of ground band only, nearest-neighbor hopping and nearest-neighbor interactions in a homogeneous system \cite{Yamamoto09}; the use of ground band only and nearest-neighbor interactions \cite{Mazzarella06}; and the use of nearest-neighbor hopping and on-site interactions in a homogeneous system \cite{Larson09}.

Limiting to the ground band of a cubic lattice, nearest-neighbor hopping (and adding the energy offset $J_{0,i,i}$), and on-site interactions, the Hamiltonian reduces to the Bose-Hubbard model \cite{Hubbard63,Fisher89}, which is:
\begin{align}
-J\hspace{-0mm}\sum_{\av{i,i'}}\ahd_{0,i}\ah_{0,i'} \hspace{-0mm}+\hspace{-0mm} \sum_i\nh_{0,i}(\vi\hspace{-0mm}-\hspace{-0mm}\mu) 
 \hspace{-0mm} +\hspace{-0mm} \frac{U}{2}\sum_i\nh_{0, i}(\nh_{0, i}\hspace{-0mm}-\hspace{-0mm}1),
\end{align}
where $\av{i,i'}$ restricts the sum to nearest neighbors $i$ and $i'$, $J \equiv J_{0,i,i'}$, and $U \equiv U_{00}$.
\section{Mean-field approximation\label{c:mfield}}
\subsection{Mean-field approach: condensate and non-condensate}
We assume that the local number of condensate atoms is either macroscopic or zero \cite{Bogoliubov47,Fetter72}, so that
the field operator, $\Psih(\br)$, can be separated into a c-number condensate component (the order parameter), $\Phi(\br)$\label{p:Phi}, and a non-condensate field operator, $\psitr$\label{p:psitr}, defined by the usual broken symmetry approach, $\Phi(\br) \equiv \av{\Psih(\br)}$,  
$\psitr \equiv \Psihr - \Phi(\br)$ so that $\av{\psitr} = 0$. 

The assumption that $\Phi(\br)$ is a c-number is inaccurate near the edges of the condensate, where the local condensate density, $\norm{\Phi(\br)}^2$, is small and just below the critical temperature, since fluctuations are important in such regions.

We expand the condensate amplitude and the non-condensate field-operator in a Wannier basis:
%
 $ \Phi(\br)\hspace{-0.3mm} =\hspace{-1mm} \sum_i z_i w_0(\br\hspace{-0.3mm}-\hspace{-0.3mm}\bR_i),\hspace{0.3mm}
  \psitr\hspace{-0.3mm} = \hspace{-1mm}\sum_{b,i} \deh_{b,i} w_b(\br\hspace{-0.3mm}-\hspace{-0.3mm}\bR_i)$.
%
where we have restricted the condensate amplitude expansion to the ground band. For an ideal gas this assumption is exact, and with interactions, the approximation is justified by our assumption that interactions are perturbative relative to the band gap energy scale.

From \eqref{e:Psih} and the orthogonality, \eqref{e:onormw}, and completeness of the Wannier functions (from the completeness of the Bloch functions), we get:
\begin{align}
 z_i \equiv \av{\ah_{0,i}}\csp\deh_{0,i}\equiv \ah_{0,i}-z_i, \deh_{b,i}\equiv\ah_{b,i} \label{e:zdeh}
\end{align}
for $b$ above the ground band. The operators $\deh_{b',i}$ satisfy standard bosonic commutation relations. The condensate density is:
\begin{align}
  \norm{\Phi(\br)}^2 = \sum_{i,i'} z_i^* z_{i'} w_0^*(\br-\bR_i) w_0(\br-\bR_{i'}),\label{e:ncdef}
\end{align}
allowing for the non-locality of the Wannier states, with condensate number:
\begin{align}
 N_c \equiv \int\dbr\,\norm{\Phi(\br)}^2 = \sum_{i} \norm{z_i}^2.\label{e:Ncintcalc}
\end{align}
For the non-condensate, we assume that the thermal coherence length is sufficiently short (long range coherence is absorbed by the condensate) that the non-condensate one-body density matrix is diagonal in lattice site indices, so that the non-condensate density is then given by:
\begin{align}
 \av{\psitdr\psitr} 
 = \sum_{b,i} \nt_{b,i} \norm{w_b(\br-\bR_i)}^2,\label{e:ntpsidef}
\end{align}
with $\nt_{b,i} \equiv \av{\dehd_{b, i}\deh_{b,i}}$. The total non-condensate population is:   
\begin{align}
  \Nt \equiv \int\dbr\,\av{\psitdr\psitr} = \sum_{b,i} \nt_{b,i}, \label{e:Ntintcalc} 
\end{align}
and we define the $b$ band non-condensate population as $\Nt_b \equiv  \sum_{i} \nt_{b,i}$\label{p:Nt}.
\subsection{HFBP Hamiltonian\label{s:quadham}}
To express the Hamiltonian in terms of the amplitudes $z_i$, and operators, $\deh_{b,i}$, we substitute \eqref{e:zdeh} into \eqref{e:Khi} \cite{epapsref}. However, the Hamiltonian still includes up to fourth powers in the operators $\deh_{b,i}$. We make a quadratic Hamiltonian simplification by making a mean-field approximation motivated by Wick's theorem \cite{Griffin96, Giorgini97b}. This is valid in the weakly-interacting regime; therefore, our work is not valid in the strongly-correlated Mott-insulator case. In a 3D cubic lattice, the Mott-insulator transition occurs for the unit-filled system when $U/6J > 5.83$ at $T=0$ \cite{Jaksch98,Oosten01}. For typical experimental parameters, the transition occurs in \Rb when $V \gtrsim 13 E_R$ (where $V=\prod_j V_j^{1/d}$), but can be $V \gtrsim 16 E_R$ for \Na \cite{Xu06} (the scattering length of \Na is smaller than \Rb, and  \cite{Xu06} used a large lattice spacing). The lattice depth for the Mott-insulator transition is increased for higher filling factors.

Making the usual HFBP approximation \cite{Griffin96,Morgan00,epapsref}, we obtain a quadratic Hamiltonian. Separating this Hamiltonian 
by the number of depletion operators $\dehd_i$ and $\deh_i$ appearing and by band:
\begin{align}
  \Kq \equiv \sum_i \left(\Kh_{0,i} + \Kh_{1,i} + \Khd_{1,i} + \sum_b \Kh_{2,b,i}\right),\label{e:Kq}
\end{align}
with:
\begin{align}
  \Kh_{0,i} &\equiv z_i^*\left(-\sum_{i'} J_{0,i,i'} \Sh{i',i}  + \vi - \mu +  \frac{U_{00}}{2}\norm{z_i}^2\right)z_i,\\
  \Kh_{1,i} &\equiv \dehd_{0,i}\left(-\sum_{i'} J_{0,i,i'}\Sh{i',i} + \vi - \mu \right.\notag\\&+\left. U_{00}\norm{z_i}^2 + 2\sum_{b} U_{0b}\nt_{b,i} \right)z_i,\\
  \Kh_{2,b,i} &\equiv \dehd_{b,i}\Lh_{b,i}\deh_{b,i} + \frac{U_{0b}}{2} \left(\dehds_{b,i}z^2_{i} + \dehs_{b,i}z_{i}^{*2}  \right),\label{e:K2bidef}
\end{align}
where:
\begin{align}
  \Lh_{b,i} &\equiv -\hspace{-1mm}\sum_{i'} \hspace{-0.5mm}J_{b,i,i'}\Sh{i',i} + \vi \hspace{-0.5mm}- \hspace{-0.5mm}\mu 
  + 2U_{0b}\norm{z_i}^2 \hspace{-0.5mm}+ 2\sum_{b'}U_{bb'}\nt_{b',i},\label{e:Ldef}
\end{align}
and $\Sh{i',i}$ is the shift operator from the site $\bR_i$ to $\bR_{i'}$, e.g. $\Sh{i',i}\deh_{b,i} = \deh_{b,i'}$.
\subsection{\gp equation}
By minimizing the energy functional $\tdiffl{\av{\Kq}}{z_i^*}=0$, using $\av{\dehd_{0,i}}=\av{\deh_{0,i}}=0$, we obtain the generalized \gp equation:
\begin{align}
&\left(-\sum_{i'} J_{0,i,i'} \Sh{i',i}  + \vi - \mu +  U_{00}\norm{z_i}^2 + 2\sum_{b} U_{0b}\nt_{b,i}\right)z_i\label{e:gengpe}\notag\\
&\hspace{5cm}= 0.
\end{align}
We note that if $z_i$ satisfies the generalized \gp equation, then the terms $\Kh_{1,i}$ and $\Khd_{1,i}$ are zero and the next contribution comes from $\Kh_{2,b,i}$.

When the interaction and trap energy is much more significant than the hopping energy, \eqref{e:gengpe} has the Thomas-Fermi solution:
\begin{align}
  \norm{z_i}^2 = \frac{1}{U_{00}}\max\left(0,\mu - \vi-  2\sum_{b} U_{0b}\nt_{b,i}\right),\label{e:TFd}
\end{align}
where $\mu$ is determined by $N=\sum_i \norm{z_i}^2 + \sum_{b,i} \nt_{b,i}$.
\subsection{Hartree-Fock}
The Hartree-Fock treatment is obtained by ignoring the terms $\dehds_{b,i}z^2_{i}$ and $\dehs_{b,i}z_{i}^{*2}$ in $\Kh_{2,b,i}$ which can then be diagonalized by a single particle transformation, setting $\deh_{b,i} = \sum\dc_j u_{b,i,j}\alh_{b,j}$ (where the symbol $\sum\dc_j$ indicates a sum over modes excluding the condensate). The operators $\alh_{b,j}$ are chosen to satisfy usual bosonic commutation relations:
\begin{align}
  \com{\alh_{b,j}, \alhd_{b',j'}} = \delta_{bb'}\delta_{jj'}\csp\com{\alh_{b,j}, \alh_{b',j'}} = 0, \label{e:alcom}
\end{align}
and the $u_{b,i,j}$ modes are an orthonormal basis, i.e. $\sum_{i} u^*_{b,i,j} u_{b,i,j'}= \delta_{jj'}$, satisfying: 
\begin{align}
  \Lh_{b,i}u_{b,i,j} = E_{b,j} u_{b,i,j}, \label{e:LuEuHF}
\end{align}
so that
%
 $ \sum_i\Kh_{2,b,i} 
  = \sum\dc_{j}  E_{b,j} \alhd_{b,j} \alh_{b,j}$. 
%
Taking the condensate to  satisfy the generalized \gp equation, we have $\Kq \rightarrow \Kh_{\mathrm{HF}}$, with:
\begin{align}
  \Kh_{\mathrm{HF}} &\equiv \sum_{i} z_i^*\left(-\sum_{i'} J_{0,i,i'} \Sh{i',i}  + \vi - \mu +  \frac{U_{00}}{2}\norm{z_i}^2\right)z_i\notag\\&+ \sum_{b,j}\dc  E_{b,j} \alhd_{b,j} \alh_{b,j}.
\end{align}
Since the Hamiltonian is diagonal in band $b$ and mode $j$, we can treat the Hartree-Fock modes as non-interacting, so that the non-condensate is given by
%
 $ \nt_{b,i} = \sum\dc_j \norm{u_{b,i,j}}^2 \nbe(E_{b,j})$, 
%
where $\nbe(E)\equiv(e^{\beta E}-1)^{-1}$.
\subsection{Quasi-particle treatment}
In general, it is desirable to go beyond the Hartree-Fock treatment when the condensate is present, to more fully include the effect of the condensate on the excitations of the system (the lattice  makes this more important, see section \ref{c:results}). To do this, we retain the terms $\dehds_{b,i}z^2_{i}$ and $\dehs_{b,i}z_{i}^{*2}$ in the Hamiltonian, which can be diagonalized using a 
quasi-particle transformation \cite{Bogoliubov47}:
\begin{align}
  \deh_{b,i} &= \sum_j\dc\left(u_{b,i,j}\alh_{b,j} + v^*_{b,i,j}\alhd_{b,j}\right),
\label{e:dehuv}
\end{align}
where we refer to the $\alh_{b,j}$ as the quasi-particle operators and the $u_{b,i,j}, v_{b,i,j}$ as the quasi-particle modes. 
We require that \eqref{e:alcom} holds, as for the Hartree-Fock case so that 
%
$\sum\dc_{j}\left(u_{b,i,j}u^*_{b,i',j} - v^*_{b,i,j}v_{b,i',j}\right) = \delta_{ii'}$, 
%
and
$\com{\deh_{b,i},\deh_{b,i'}}  
      = \sum\dc_{j} \left(\pst{u}_{b,i,j}\st{v}_{b,i',j} -\st{v}_{b,i,j}\pst{u}_{b,i',j}\right) = 0$.
The quasi-particle modes are normalized according to 
%
  $\sum_{i} \left(\norm{u_{b,i,j}}^2 - \norm{v_{b,i,j}}^2\right) = 1$. 
%
We choose the modes to satisfy the Bogoliubov-de Gennes equations:
\begin{align}
  \Lh_{b,i}u_{b,i,j} + U_{0b}z_i^2 v_{b,i,j} &= E_{b,j} u_{b,i,j},\label{e:Lhu}\\
  \Lh_{b,i}v_{b,i,j} + U_{0b}z^{*2}_i u_{b,i,j} &= -E_{b,j} v_{b,i,j}.\label{e:Lhv}
\end{align}
The Hamiltonian \Kq is diagonal with these solutions \cite{epapsref}:
\begin{align}
  \Kq &= 
 \sum_{i} z_i^*\left(-\sum_{i'} J_{0,i,i'} \Sh{i',i}  + \vi - \mu +  \frac{U_{00}}{2}\norm{z_i}^2\right)z_i\notag\\ 
  &+ \sum_{b,j}\dc  E_{b,j}\left(\alhd_{b,j}\alh_{b,j} -  \sum_i \norm{v_{b,i,j}}^2 \right)\label{e:Khdiag},
\end{align}
and we can treat the quasi-particles as non-interacting which leads to: 
\begin{align}
  \nt_{b,i}       &= \sum_{j}\dc\left(\norm{u_{b,i,j}}^2 + \norm{v_{b,i,j}}^2\right)\nbe(E_{b,j}) + \norm{v_{b,i,j}}^2.\label{e:ntbi}
\end{align}
The references \cite{Morgan99,Morgan00} explain that, for a general potential, 
\eqref{e:Lhu} and \eqref{e:Lhv} give quasi-particle functions which are orthogonal to the condensate only in a generalized sense, $\sum_i z_i^* u_{b,i,j} + z_i v_{b,i,j} = 0$. To be orthogonal in the sense $\sum_i z_i^* u_{b,i,j} = \sum_i z_i v_{b,i,j} = 0$, adjustments are required, e.g. $E_{0,j} u_{0,i,j}$ is replaced by \cite{Burnett02,Rey03}:
\begin{align}
E_{0,j} u_{0,i,j} + U_{00}\sum_i \norm{z_i}^2\left(z_i^* u_{0,i,j} - z_i v_{0,i,j}\right)z_i.
\end{align}
We do not follow this approach since, in our LDA solution below, we approximate by using an orthogonal Bloch form for the modes.
\section{Local density approximation\label{c:lda}}
The LDA has been extensively used for (non-lattice) harmonically trapped Bose gases. The essence of this approximation is the replacement $-\hbar^2\nabla^2/2m \rightarrow p^2/2m$ in the Hamiltonian with $\br$ and $\bp$ treated as classical variables. The extension of this approach to the lattice case is made by the replacement $\Hl \rightarrow K_b(\bk)$ where $\bk$ is the quasi-momentum, $b$ the quantized band index and $K_b(\bk)$ the Bloch spectrum. In what follows, we present our assumptions in making this replacement.
\subsection{Bloch approximation}
We set $j$ to be the quasi-momentum, $\bk$, and make the LDA by seeking solutions where $u$ and $v$ have the Bloch form:
\begin{align}
  u_{b,i',\bk} = e^{\ii \bk \cdot (\bR_{i'}-\bR_{i})}u_{b,i,\bk}\csp v_{b,i',\bk} = e^{\ii \bk \cdot (\bR_{i'}-\bR_{i})}v_{b,i,\bk}.\label{e:uvbkbloch}
\end{align}
This assumption is exact for the \unif case, and we justify it in general by comparing the non-interacting density of states obtained using this approximation to the numerical diagonalization of the full combined harmonic lattice problem in section \ref{s:traplat}.

To make progress, it is useful to consider the Bloch waves,
$\psi_{b,\bk}(\br)$, of \Hl:
\begin{align}
  \Hl\psi_{b,\bk}(\br) = K_b(\bk)\psi_{b,\bk}(\br),\label{e:Hnilatt}
\end{align}
which serves to define the energy, $K_b(\bk)$. We find from \eqref{e:uvbkbloch} and \eqref{e:enkF} that 
%
$-\sum_{i'}
J_{b,i,i'} u_{b,i',\bk} 
      =  K_b(\bk)u_{b,i,\bk}$, so that: 
%
%
\begin{align}
  &\Lh_{b,i}u_{b,i,\bk} \notag\\&= \left(K_b(\bk) + \vi - \mu + 2U_{0b}\norm{z_i}^2 + 2\sum_{b'}U_{bb'}\nt_{b',i}\right) u_{b,i,\bk}.\label{e:Lhuk}
\end{align}
\subsection{Envelope functions \label{s:envelope}}
We define a function $\ntd{b}(\br)$\label{p:ntbr} which is a proxy with the continuous variable $\br$ for the number of non-condensate atoms per site: $\ntd{b}(\bR_i) = \nt_{b,i}$. Introducing this envelope function greatly simplifies our formalism by allowing us to use continuous functions to exploit the symmetry of \vt, which is broken on short length scales by the lattice. Then, for a sufficiently small lattice spacing:
\begin{align}
  \frac{1}{a^d}\int\dbr\, \ntd{b}(\br) \approx \sum_i \ntd{b}(\bR_i) = \sum_i \nt_{b,i} = \Nt_b, \label{e:Ntintnt}
\end{align}
where $a^d$ is the volume of a unit cell of the optical lattice. 
Similarly, we define the condensate mode envelope $z(\br)$, where $z(\bR_i) = z_i$ and 
$\ncd(\br) \equiv \norm{z(\br)}^2$, 
so that: 
\begin{align}
  \frac{1}{a^d}\int\dbr\, \ncd(\br) \approx \sum_i \norm{z(\bR_i)}^2 = \sum_i \norm{z_i}^2 = N_c.
\end{align}
We also define the envelope functions $u_b(\bk,\br)$ and $v_b(\bk,\br)$, with $u_b(\bk,\bR_i) = u_{b,i,\bk}$ and $v_b(\bk,\bR_i) = v_{b,i,\bk}$, and from \eqref{e:Lhuk} we have $\Lh_{b,i} \rightarrow \Lnh_b(\bk,\br)$ where:
\begin{align}
   \Lnh_b(\bk,\br) &=K_b(\bk) + \vt - \mu \notag\\&+ 2U_{0b}\ncd(\br) +    2\sum_{b'} U_{bb'}\ntd{b'}(\br).\label{e:Lbkr}
\end{align}
Envelope functions represent the discrete functions and do not contain the fast Wannier state variation. However, apart from exceptional imaging techniques \cite{Gericke08}, normal optical imaging techniques would not distinguish density variation at the order of one site. If we require the detailed spatial density, rather than just site occupation, once we have the envelope functions, we can calculate
$\norm{\Phi(\br)}^2 = \sum_{i,i'} z^*(\bR_i)z(\bR_{i'}) w_0^*(\br-\bR_i) w_0(\br-\bR_{i'})$
from \eqref{e:ncdef}
and
$\av{\psitdr\psitr} = \sum_{b,i} \ntd{b}(\bR_i)\norm{w_b(\br-\bR_i)}^2$
from \eqref{e:ntpsidef}.
\subsection{Bogoliubov spectrum}
Making use of the envelope functions from the previous section, the Bogoliubov-de Gennes equations, \eqref{e:Lhu} and \eqref{e:Lhv}, take the algebraic form:
\begin{align}
  \begin{bmatrix}
  \Lnh_b(\bk,\br) & U_{0b} z^2(\br)\\
  -U_{0b} z^{*2}(\br)  & -\Lnh_b(\bk,\br)
  \end{bmatrix}
  \begin{bmatrix}
  u_b(\bk,\br)\\
  v_b(\bk,\br)
  \end{bmatrix}
  = E_b(\bk,\br)
  \begin{bmatrix}
  u_b(\bk,\br) \\
  v_b(\bk,\br)
  \end{bmatrix}
  . \label{e:scuv}
\end{align}
Solving the characteristic equation yields:
\begin{align}
E_b(\bk,\br) &= \sqrt{\Lnh^2_b(\bk,\br)-\left[U_{0b}\ncd(\br)\right]^2}.
                             \label{e:sce}
\end{align}
From \eqref{e:scuv}, choosing the normalization condition $\norm{u_b(\bk,\br)}^2 - \norm{v_b(\bk,\br)}^2 = 1$ (as in \cite{Giorgini97b} for the no lattice case) we have:
\begin{align}
\norm{u_b(\bk,\br) }^2
&= \frac{\Lnh_b(\bk,\br) + E_b(\bk,\br)}{2E_b(\bk,\br)},\\
\norm{v_b(\bk,\br) }^2
&= \frac{\Lnh_b(\bk,\br) - E_b(\bk,\br)}{2E_b(\bk,\br)}.\label{e:vsqkr}
\end{align}
Setting $v_b(\bk,\br)=0$, we find $\norm{u_b(\bk,\br)}^2 = 1$ and $E_b(\bk,\br) = \Lnh_b(\bk,\br)$, yielding the LDA envelope form of the Hartree-Fock solution \eqref{e:LuEuHF}.  

It has been stated that the Thomas-Fermi approximation is necessary to be consistent with the LDA \cite{Reidl99}. We use the Thomas-Fermi solution for all of our interacting calculations, which we restate using the envelope functions, starting from \eqref{e:TFd} to find: 
\begin{align}
  \ncd(\br) = \frac{1}{U_{00}}\max\left[0,\mu - \vt-  2\sum_{b} U_{0b}\ntd{b}(\br)\right].\label{e:ncdTF}
\end{align}
For the non-condensate, using \eqref{e:ntbi} and the envelope functions we have (\bz is the first Brillouin zone):
\begin{align}
  \ntd{b}&(\br) 
   = \fracb{a}{2\pi}^d\int_{\bz} \dbk  \left\{\right.\notag\\&\left.\left[\norm{u_b(\bk,\br)}^2+ \norm{v_b(\bk,\br)}^2\right]\nbe[E_b(\bk,\br)] + \norm{v_b(\bk,\br)}^2 \right\} .\label{e:nbdintk}
\end{align}
From \eqref{e:sce}, if $\nc(\br)$ is zero (e.g. above $T_c$ or outside the Thomas-Fermi radius), we have the Hartree-Fock result. Otherwise, for the ground band, from \eqref{e:ncdTF}:
\begin{align}
 \Lnh_0(\bk,\br) &= K_0(\bk) + U_{00}\ncd(\br),\\
  E_0(\bk,\br) &= \sqrt{K^2_0(\bk) + 2K_0(\bk) U_{00}\ncd(\br)},
\end{align}
which is a useful simplification, and is automatically self-consistent with $\ncd(\br)$.

If we rearrange the equation for the non-condensate envelope \eqref{e:nbdintk}, we obtain:
\begin{align}
 & \ntd{0}(\br) = \fracb{a}{2\pi}^d \int_{\bz} \dbk  
  \left\{\frac{ K_0(\bk) + U_{00}\ncd(\br)}{E_0(\bk,\br)}\nbe[E_0(\bk,\br)] \bsplit\hspace{1cm}+ \frac{K_0(\bk) + U_{00}\ncd(\br) - E_0(\bk,\br)}{2E_0(\bk,\br)} \right\}\notag\\
         &= \fracb{a}{2\pi}^d \hspace{-1mm}\int_{\bz}\hspace{-1mm} \frac{\dbk}{2}\hspace{-1mm}
  \left\{\frac{ K_0(\bk)\hspace{-0.8mm} +\hspace{-0.8mm} U_{00}\ncd(\br)}{E_0(\bk,\br)}
  \coth\hspace{-1mm}\left[\frac{\beta E_0(\bk,\br)}{2}\right]\hspace{-1mm}- \hspace{-0.7mm} 1\right\},\label{e:nttfsimple}
\end{align}
If $K_0(\bk)$ is restricted to nearest-neighbor hopping, then this result is consistent with that given by Duan and co-workers \cite{Lin08}. We note that they do not make the envelope approximation (the discrete LDA sum in their Eqn.~(15) should have been divided by the number of sites). Additionally, their theory is restricted to the ground band, and is stated for a cubic lattice and a spherical harmonic trap.
\section{Density of states\label{c:dos}}
The theory we develop relies on detailed knowledge of the density of states of the \unif lattice.
\subsection{Definition and usage}
By `density of states', we refer to the per-site density of states for the non-interacting, \unif  lattice which we  define as \cite{Ashcroft76}:
\begin{align}
      g_b(K)  &\equiv \frac{1}{(2\pi)^d}\int_{\bz}\dbk\,\delta[K - K_b(\bk)],\label{e:gdosdef}
\end{align}
where we take $K_b(\bk)$ from its definition \eqref{e:Hnilatt}.
When an integrand depends on $\bk$ only through $K_b(\bk)$ we can change variables to $K = K_b(\bk)$ since we then have, for any function $Q_b[K_b(\bk),\br]$:
\begin{align}
  \intinf \diff K\,g_b(K) Q_b(K,\br) 
  &= \frac{1}{(2\pi)^d}\int_{\bz} \dbk\, Q_b[K_b(\bk),\br]. \label{e:doscvar}
\end{align}
Applying this to \eqref{e:nbdintk}:
\begin{align}
  \ntd{b}(\br) &= a^d\intinf \diff K \,g_b(K)  \left\{
  \frac{\Lnh_b(K,\br)}{E_b(K,\br)}
  \nbe[E_b(K,\br)]   
  \right.\notag\\&\left.
+ 
  \frac{\Lnh_b(K,\br) - E_b(K,\br)}{2E_b(K,\br)}
   \right\}.\label{e:ntbhfbp}
\end{align}
We emphasize that this is making no additional approximations. Similarly, in the Hartree-Fock approach, or above the critical temperature, 
%
 $ \ntd{b}(\br) = a^d\intinf \diff K \,g_b(K)\,\nbe[E_b(K,\br)]$. 

\begin{figure}[t] 
  \centering   
  {\small\includegraphics{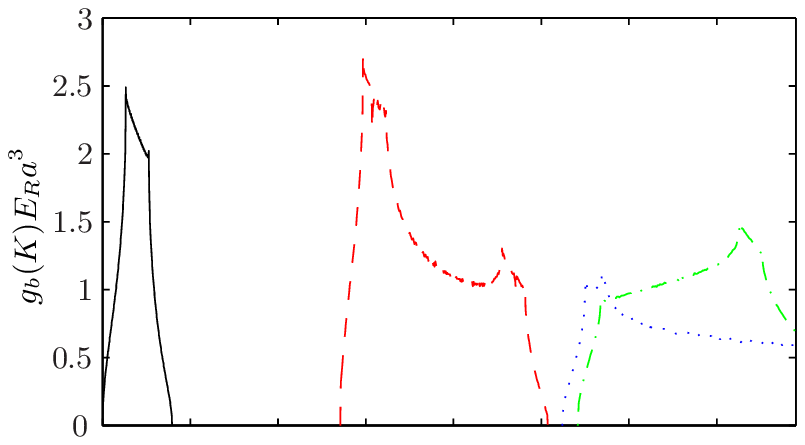}  \put(-25,105){(a)}}
  {\small\includegraphics[trim=0 2 0 5,clip=true]{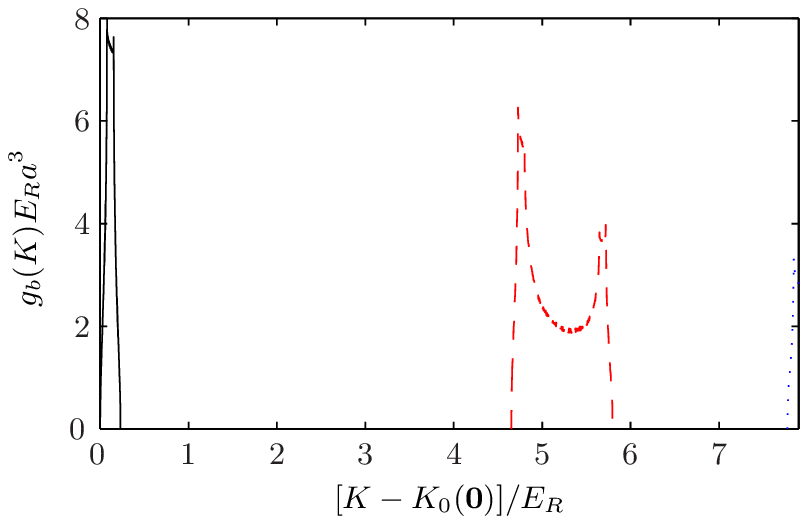}  \put(-25,125){(b)}}
  \caption{(Color online) Density of states for the 3D cubic lattice: $b=0$ (black solid curve), 001 (red dashed curve, the integers specify the components $b_x,b_y,b_z$), 011 (green dashed-dotted curve),  002 (blue dotted curve) for (a) $V=5E_R$ and (b) $V=10E_R$}
  \label{f:dos15}
\end{figure}

To calculate the density of states, we first need the energy
dispersion, $K_b(\bk)$, which is easy if the lattice potential is separable (the well-studied Mathieu's equation \cite{Abramowitz70,McLachlan64,Muller-Seydlitz97}), but separability is not required. We numerically calculate the density of states and show the results in Fig.~\ref{f:dos15}.
\subsection{Limiting results for the \unif lattice\label{s:doslimit}}
\subsubsection{Tight binding}
From \eqref{e:enkfouriercos}, the dispersion can be written as a Fourier cosine series, with the hopping matrix elements as coefficients:
\begin{align}
  K_b(\bk) = -\sum_{j=1}^d \left[ J^0_{b_j,j} + 2\sum_{l>0} J^l_{b_j,j} \cos(l k_j a_j) \right],\label{e:KcosF}
\end{align}
for a separable lattice, where we define the band $b$ hopping between neighbors $l$ sites apart in axial direction $j$ to be $J^l_{b,j}$ (e.g. $J^l_{b,y} = J_{b,000,0l0}$ and, for the cubic lattice, $J=J^1_{0,j}$)\afn[.]{When we use this notation, we are implicitly assuming that the energy spectrum is invariant under inversion of quasi-momentum, in view of \eqref{e:jni}.}

\begin{figure}[t] 
  \centering   
  {\small\includegraphics{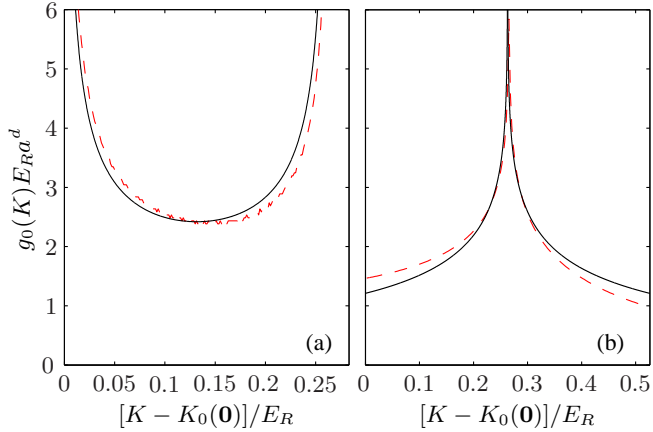}}
  \put(-135,30){(a)}
  \put(-25,30){(b)}
  \caption{(Color online) Tight-binding (black solid curve) and actual (red dashed curve) density of states for $V=5E_R$ for (a) 1D and (b) the 2D square lattice}
  \label{f:dostb12d}
\end{figure}

\begin{figure}[t] 
  \centering   
  {\small\includegraphics{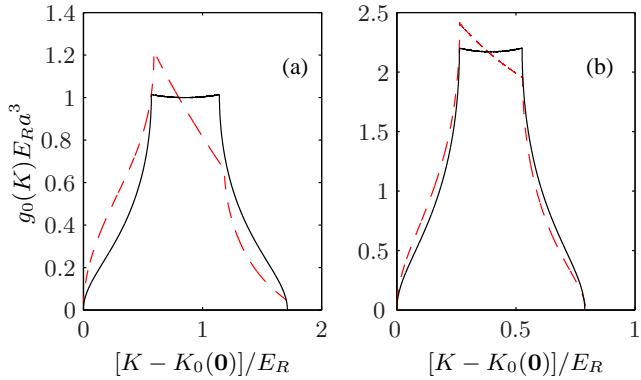}}
  \put(-145,115){(a)}
  \put(-30,115){(b)}
  \caption{(Color online) Tight-binding (black solid curve) and actual (red dashed curve) 3D cubic-lattice ground-band density of states for (a) $V=2E_R$ and (b) $V=5E_R$}
  \label{f:dostb3d}
\end{figure}

In the tight-binding limit, beyond nearest-neighbor hopping is ignored (for the importance of beyond nearest-neighbor hopping, see also section \ref{s:bnnhresult} and appendix \ref{a:appwanhop}). In 1D, the density of states is then, from \eqref{e:gdosdef}, 
   $g_0(K) 
     = 1/\left\{ 2\pi a \norm{J^1_{0,j}} \sqrt{1-\left[(K+J^0_{0,j})/2J^1_{0,j}\right]^2}\right\}$, 
which has infinite van Hove singularities at the maximum and minimum energies of the band, which can also be seen from the zero derivative in \eqref{e:KcosF}. In 2D, the square-lattice density of states\afn{By convolution we can express it as a complete elliptic integral of the first kind as
 $g_0(K)\hspace{-0.7mm} =\hspace{-0.7mm} K\left\{1\hspace{-0.7mm}-\hspace{-0.7mm}\left[(K\hspace{-0.7mm}+\hspace{-0.7mm}2J^0_{0,j})/2J^1_{0,j}\right]^2/4\right\}/\left(2\pi^2 a^2\norm{J^1_{0,j}}\right)$} has an infinite van Hove singularity at the band center and non-zero density at the band edges. The density of states for 1D and 2D are shown in Fig.~\ref{f:dostb12d}.
In 3D, we compare the tight-binding density of states to the actual density of states in Fig.~\ref{f:dostb3d} for the cubic-lattice ground band. For $V \gtrsim 5E_R$, the effect of beyond nearest-neighbors is much reduced, except for very low energies.
\subsubsection{Effective mass\label{s:dosemass}}
If, at the minimum energy of a band ($\Kbm = K_b(\bk_0)$), we have $\nabla K_b(\bk_0) = \bzero$, then  from the quadratic Taylor series, we get the effective mass approximation 
$K_b(\bk) \approx K_b(\bk_0) + \sum_j \hbar^2 k_j^2/2m^*_j$ 
where $m_j^*$ is the effective mass at $\bk_0$ in direction $j$, 
$
   \fracl{1}{m_j^*} \equiv \pdiffal{^2 K_{b}(\bk)}{k_j^2}{\bk=\bk_0}/{\hbar^2}
$
\cite{Ashcroft76}. If, due to the second derivative test, we have $m^*_j>0$ for all $j$ and assuming that the effective mass approximation applies for all $K$ in some region near $\Kbm$ (for excited bands and deep lattices, there is only a small region around $\bk_0$ for which this is a good approximation), then for that region of $K$, from \eqref{e:gdosdef}:
\begin{equation}
 g_b(K) = \frac{\max \left(K-\Kbm,0\right)^{d/2-1}}{\Gamma(d/2)(2\pi)^{d/2} \left(\hbar^2/m^*\right)^{d/2}} 
 , \label{e:gkeffmass}
\end{equation}
where $m^{*} \equiv {\prod_j m^*_j}^{1/d}$. We note this shows that the van Hove singularities at the minimum energy are qualitatively the same for the effective-mass assumption as for the tight-binding assumption: infinite in 1D, a finite jump in 2D and an infinite derivative in 3D.\vspace{-0.4cm}
\subsubsection{High energies\label{s:highe}}
\begin{figure}[t] 
  \centering   
  {\small\includegraphics{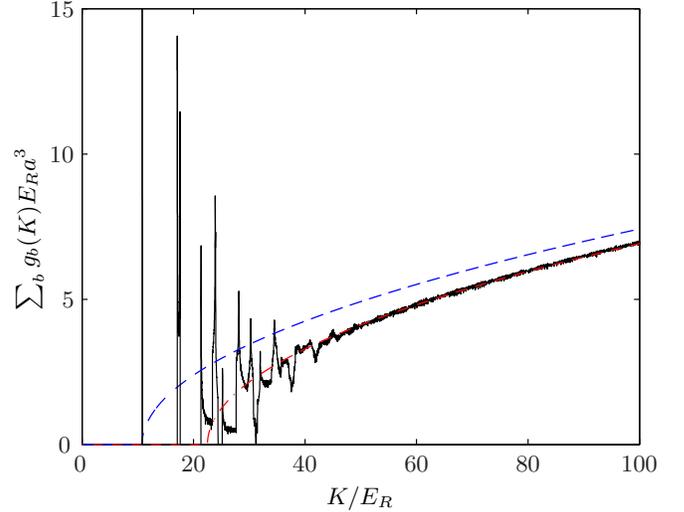}}
  \caption{(Color online) 3D cubic-lattice density of states for $V=15E_R$ (black solid curve), the free-particle density of states shifted by the minimum energy eigenvalue ($\frac{\pi}{4}\sqrt{[K-K_0(\bzero)]/E_R}$, blue dashed curve) and by the spatially averaged energy of the lattice ($\frac{\pi}{4}\sqrt{(K-\half\sj V_j)/E_R}$, red dashed-dotted curve).  }
  \label{f:doshighe}
\end{figure}

For high energies, $K \gg \sj V_j$, the most significant effect of the lattice on the density of states is the spatially averaged energy of the lattice potential, $\half \sj V_j$ as shown in Fig.~\ref{f:doshighe}.
\subsection{Limiting results for the combined harmonic lattice\label{s:traplat}}
In this section, we consider the LDA density of states for the combined harmonic trap and optical lattice potential (some features of the combined harmonic lattice density of states in the 1D tight-binding case, and the 2D case, numerically, are discussed  in \cite{Hooley04}). We introduce the LDA density of states for comparison with the full numerical diagonalization as justification of the validity of the LDA approach.

For the harmonically trapped case, in the non-interacting LDA, when we wish to calculate some function, $Q[K_b(\bk) + \vt]$ of the energy, such as the total number of non-condensate atoms \eqref{e:Ntintnt} and \eqref{e:nbdintk}, we have:
\begin{align}
    &\frac{1}{(2\pi)^d} \sum_b \int\dbr\int_\bz\dbk\,Q[K_b(\bk) + \vt]\notag\\
 &= \int\diff E\,Q(E)\,\ga(E),
\end{align}
from \eqref{e:doscvar} where $\ga(E)$ is given by the convolution:\enlargethispage{1cm}
\begin{align}
  \ga(E)  & \equiv \frac{1}{(2\pi)^d}\sum_b\int\dbr \int_{\bz}\dbk \,\delta[E-K_b(\bk)-\vt]\notag\\
 & =\sum_b\int_0^E\diff \vtn\, \gt(\vtn)g_b(E-\vtn),\label{e:gall}
\end{align}
with:
\begin{align}
      \gt(\vtn)  &\equiv \int\dbr\,\delta[\vtn - \vt] 
      = \frac{(2\pi)^{d/2}}{\Gamma(d/2) \left(m\omega^2\right)^{d/2}}  \vtn^{d/2-1}. \label{e:gtrap}
\end{align}
Since the combined density of states, $\ga(E)$, has a rich structure, we consider what we expect at various energies. 
In a region where the effective-mass approximation, \eqref{e:gkeffmass}, applies, the contribution to $\ga(E)$ from band $b$ is:
\begin{align}
\frac{1}{(d-1)! (\hbar\omega^*)^d}\left(E-\Kbm\right)^{d-1},\label{e:emassgall}
\end{align}
where the effective trap frequencies are defined by:
\begin{align}
  \omega_j^* \equiv \sqrt{\frac{m}{m_j^*}}\omega_j, \label{e:omegastar}
\end{align}
as in \cite{Rey05} and $\omega^* = {\prod_j \omega_j^*}^{1/d}$\label{p:omegastar}. We therefore expect the initial contribution from each band (just after $\Kbm$) to the combined density of states to scale like a harmonically-trapped particle, with power ${d-1}$.

If we assume that the bands are rectangular with width $W_b$\label{p:Wb} and minimum energy $\Kbm$, so that $g_b(K) = 1/(W_b a^d)$ for $\Kbm < K < \Kbm + W_b$ and $g_b(K)=0$ otherwise, then:
\begin{align}
  \ga&(E) \approx \frac{2(2\pi)^{d/2}}{d\Gamma(d/2) \left(m\omega^2a^2\right)^{d/2}} 
  \sum_b \left[\phantom{\frac{\max\left(E-\Kbm,\hspace{-0.1mm}0\right)^{d/2} \hspace{-1.5mm} - \max\left(E-\Kbm-W_b,\hspace{-0.1mm}0\right)^{d/2}}{W_b}}\bsplit
  \hspace{-0.4cm}\frac{\max\left(E-\Kbm,\hspace{-0.1mm}0\right)^{d/2} \hspace{-1.5mm} - \max\left(E-\Kbm-W_b,\hspace{-0.1mm}0\right)^{d/2}}{W_b}\right]\label{e:gallpd2}\\
  &\approx \frac{(2\pi)^{d/2}}{\Gamma(d/2) \left(m\omega^2a^2\right)^{d/2}}
   \sum_b \left(E-\Kbm-\half[W_b]\right)^{d/2-1}\notag\\
  &=\frac{1}{a^d}\sum_b \gt\left(E-\Kbm-\half[W_b]\right),\label{e:gallsumg}
\end{align}
for $E \gg \Kbm + W_b$, using \eqref{e:gall} and \eqref{e:gtrap}. So, we expect the eventual contribution of the band to the combined density of states  (far after $\Kbm+W_b$) to scale like the trap, with power $d/2-1$. The high-energy contribution is therefore like the density of states for a particle in a harmonic trap with no kinetic energy, we call this the `trap-only' region.

For energies beyond the effective-mass region, but with $\Kbm<E<\Kbm+W_b$, the combined density of states depends on the detailed structure of the band $g_b(K)$ with an approximation given by \eqref{e:gallpd2}\afn[.]{For\label{fn:lowdpoor} $\Kbm<E<\Kbm+W_b$ the rectangular assumption implies that the contribution to $\ga(E)$ from band $b$ is proportional to $(E-\Kbm)^{d/2}$. For 3D, this is a blend between the effective-mass (power $d-1$) behavior near the start of the band and the trap-only (power $d/2-1$) behavior far after the band. For lower dimensions, the rectangular assumption is poor from Fig. \ref{f:dostb12d}.}

So, the initial contribution from the band is effective-mass like and the high-energy contribution from the band is trap-only like. We estimate the crossover point between these two regimes by equating the single-band contribution from equations \eqref{e:emassgall} and \eqref{e:gallsumg}. In 3D there is no intersection for the first excited bands for $V \gtrsim 5E_R$ and, for the ground band:
\begin{align}
  \Ele - \Kbm[0] &=\half[W_0] + \frac{1}{128\pi^2}\fracb{m^*a^2}{\hbar^2}^{3}\left(\Ele-\Kbm[0]\right)^4.
\end{align}
Using the tight-binding approximations \eqref{e:widthj} and 
 $ \fracl{m}{m_j^*} \approx \fracl{\pi^2 J^1_{0,j}}{\erj}$ \cite{Rey05} (where $E_{R,j} \equiv h^2/2m\lambda_j^2$)
,
 for the cubic lattice and assuming that the cross over is near the middle of the band $\Ele-\Kbm[0]\approx W_0/2$:
\begin{align}
  \Ele - \Kbm[0] &\approx  \left(\frac{1}{2} +   \frac{27}{256 \pi^2}\right) W_0 
   \approx  0.51 \,W_0, \label{e:ecross}
\end{align}
as shown in Fig. \ref{f:comdos}. This result has the same scaling, but is slightly lower than  $ \Ele - \Kbm[0]\approx 0.86 \,W_0 $, given in \cite{Blakie07a}. 

For high energies, once there have been many bands, we consider the assumption that the bands start at the free-particle positions, adjusted by the average energy of the lattice (as shown in Fig. \ref{f:doshighe}),
$\Kbm = \sj\left(\half V_j + \hbar^2\pi^2b_j^2/2ma_j^2\right)$. We keep the other assumptions leading to \eqref{e:gallsumg} and approximate the sum in \eqref{e:gallsumg} by an integral over the region of bands $b$ such that $0<\Kbm<E$, 
then we recover the density of states for a trap with no lattice (\eqref{e:emassgall} with $m=m^*$). Evaluating this integral in band space, we find: 
\begin{align}
  \ga(E) \approx \frac{1}{(d-1)! (\hbar\omega)^d}\left(E -\half \sj V_j\right)^{d-1},\label{e:dosallhe}
\end{align}
so, the eventual contribution of all bands has power $d-1$, like the density of states of a harmonically-trapped particle.
\subsection{Comparative results}
\begin{figure}[!t] 
  \centering   
  {\hspace{-6mm}\small\includegraphics{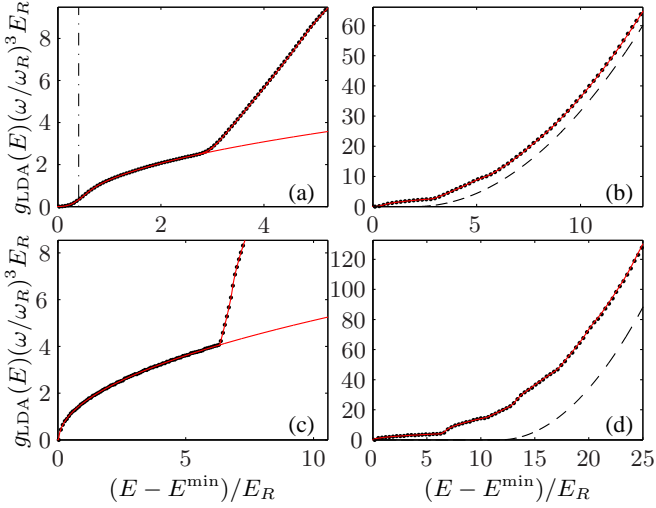}}
  \put(-140,115){(a)}
  \put(-20,115){(b)}
  \put(-140,27){(c)}
  \put(-20,27){(d)}
  \caption{(Color online) 3D combined harmonic cubic-lattice density of states for (a,b) $V=5E_R$  and (c,d) $V=15E_R$ from the full diagonalization (black dotted curve), LDA (red solid curve) and $\Ele$ (dashed-dotted curve). Shown are [(a),(c)] the ground to the first excited bands with the LDA ground band (lower red solid line) for reference,
    [(b),(d)] many bands and the high-energy  approximation \eqref{e:dosallhe} (dashed curve).
The LDA is so good that it is obscured by the full diagonalization results in all cases.  }\vspace{-0.3cm}
  \label{f:comdos}
\end{figure}

We compare the density of states obtained from the full diagonalization of $\Hl + \vt$ (see \cite{Blakie07a}) to the LDA density of states in Fig.~\ref{f:comdos}. For the low energy LDA results, we also show the contribution from the ground band. We plot the product $\ga(E)\omega^d$, since, for the LDA case, $\ga(E)\omega^d$ is independent of $\omega$ from \eqref{e:gtrap}. For the full diagonalization, we can see no dependence of the full density of states multiplied by $\omega^3$ for varying $\omega$ apart from granularity due to the few  discrete energies for large $\omega$ at low energy. \enlargethispage{1cm}
The LDA results show excellent agreement with the full diagonalization. We note that the approximation \eqref{e:dosallhe} becomes valid in the $V=15E_R$ case only for $E > E^{\mathrm{min}}+40E_R$, beyond the region of this plot. The effective-mass region is not visible on the plot for $V=15E_R$ due to the scale.
\section{Numerical implementation\label{c:numericalimpl}}
\subsection{\Unif density of states\label{s:numdos}}
We find the \unif energies, $K_b(\bk)$, from the non-interacting Bloch solutions to find the density of states, by diagonalizing the tri-diagonal (since the lattice potential is sinusoidal) Hamiltonian, \Hl, in momentum space \cite{Ashcroft76}. We calculate the density of states by binning the energies. \vspace{-0.6cm}
\subsection{Scaled units}
From \eqref{e:ncdTF} and \eqref{e:nbdintk}, 
$\ntd{b}(\br)$ and $\ncd(\br)$ depend on $\br$ only through $\vt = \half m(\omega_x^2 x^2 + \omega_y^2 y^2 + \omega_z^2 z^2 )$, so we define the scaled co-ordinates 
$\xb = x\omega_x/\omega, \yb = y\omega_y/\omega, \zb = z\omega_z/\omega, \rb^2 = \xb^2 + \yb^2 + \zb^2$\label{p:rb} so that
$\vt[\rb] = \half m \omega^2 \rb^2$ and $\diff \xb \diff \yb \diff \zb = \diff x \diff y \diff z$. Our formulae then become:
{\allowdisplaybreaks
\begin{align}
  \ncd(\rb)   = &\frac{1}{U_{00}}\max\left[0,\mu - \vt[\rb]-  2\sum_{b} U_{0b}\ntd{b}(\rb)\right],\label{e:nctrapu}\\
    \Lnh_b(K,\rb) &=K + \vt[\rb] \hspace{-0.8mm}-\hspace{-0.8mm} \mu \hspace{-0.8mm}+ \hspace{-0.8mm}2U_{0b}\ncd(\rb) \hspace{-0.8mm}+\hspace{-0.8mm} 2\sum_{b'} U_{bb'}\ntd{b'}(\rb),\label{e:Lnhtrapu}\\
    E_b(K,\rb) &= \sqrt{\Lnh^2_b(K,\rb)-\left[U_{0b}\ncd(\rb)\right]^2},\label{e:Ebtrapu}\\
  \ntd{b}(\rb) &= a^d\intinf \diff K \,g_b(K)  \left\{\frac{\Lnh_b(K,\rb)}{E_b(K,\rb)}\nbe[E_b(K,\rb)] \right.\notag\\&\left.+ \frac{\Lnh_b(K,\rb) - E_b(K,\rb)}{2E_b(K,\rb)} \right\}.\label{e:ntdbtrapu}
\end{align}
}
We can then calculate the total number using:
\begin{align}
  N_c &= \frac{2\pi^{d/2}}{\Gamma(d/2)a^d}\intzinf \diff\rb\,\rb^{d-1}\ncd(\rb),\label{e:Nc}\\
  \Nt_b &= \frac{2\pi^{d/2}}{\Gamma(d/2)a^d}\intzinf \diff\rb\,\rb^{d-1}\nt_b(\rb),\label{e:Ntb}
\end{align}
which is now a problem in the two dimensions $K$ and $\rb$, and is fundamental to our development of an efficient numerical algorithm.\vspace{-0.3cm}
\subsection{Interaction parameters}
We calculate the 1D Wannier functions and use their separability (from the separability of the Bloch functions) to get the interaction coefficients. For the cubic lattice in 3D, the densities of the three bands $001, 010$ and $100$ must be equal, i.e. $\ntd{001}(\rb) = \ntd{010}(\rb) = \ntd{100}(\rb)$. Thus we can use this symmetry to simplify our calculation of higher bands. For a given one of these bands, $\frac{1}{3}$ of the atomic population is in the same band and $\frac{2}{3}$ is in one of the other first excited bands so that:
\begin{align}
 & U_{001,001}\ntd{001}(\rb) + U_{001,010}\ntd{010}(\rb) + U_{001,100}\ntd{100}(\rb)\notag\\
&= \left(U_{001,001} + 2U_{001,010}\right)
\frac{\ntd{001}(\rb) + \ntd{010}(\rb) + \ntd{100}(\rb)}{3},
\end{align}
since $U_{001,010} = U_{001,100}$. We therefore treat the three excited bands together and use $\left(U_{001,001} + 2U_{001,010}\right)/3$ for their self-interaction parameter.
\subsection{Procedure\label{s:proc}}
We fix the parameters $N,V_j,a_j,a_s, \omega_j$ and $m$ throughout the entire calculation. For the cubic lattice, we calculate the density of states $g_b(K)$ and the interaction parameters $U_{bb'}$ once for each $V$ and use them for any cubic-lattice calculation. For the non-cubic lattice, we calculate the density of states and interaction parameters for each case.

\begin{figure}[H] 
\includegraphics[trim=355 0 85 110,clip=true]{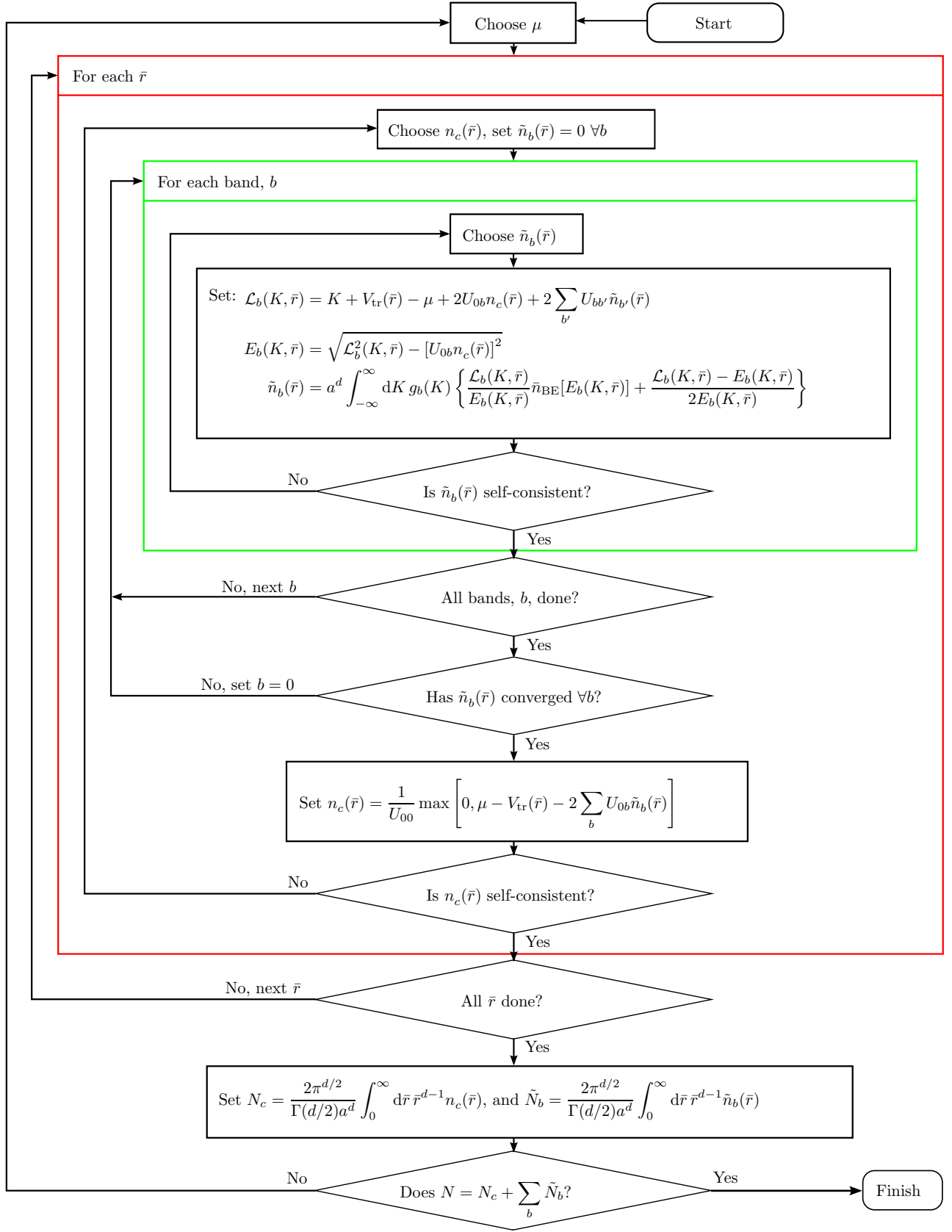}
  \caption{Procedure for LDA calculation}
  \label{f:flow}
\end{figure}

We solve \eqref{e:nctrapu}--\eqref{e:ntdbtrapu} self-consistently, finding $\mu$ so that $N=\Nc+\sum_b\Nt_b$ from \eqref{e:Nc} and \eqref{e:Ntb}. We present our algorithm for doing this in Fig.~\ref{f:flow}.
\clearpage
\clearpage
We note that, once we have a choice for the chemical potential, the calculation is completely local. Therefore, in contrast to the \gp equation approach of \cite{Giorgini97b}, we do not check the target for the total number $N$ until the calculations at every site are self-consistent. 

For the ground band we use the simplification \eqref{e:nttfsimple}, with scaled units and the density of states (this is not shown in Fig.~\ref{f:flow}).

For the \unif lattice, we use almost the same calculation, with $\vt$ set to zero, and use only one spatial point, $\rb$. However, due to the importance of the low energy states in that case, we make the substitution
$u^4=K$ and use $\int\diff K  \rightarrow \int 4u^3\diff u$ so that the integrand isn't divergent.
\subsection{Finite-size effect\label{s:fseffectimpl}}
For the non-interacting gas in a combined harmonic lattice, we allow for the effect of a positive chemical potential at condensation, equal to the minimum energy 
%
 $ \mufs \equiv \frac{d}{2}\hwb^* $, 
%
where $\omega_j^*$ are the effective trapping frequencies, defined in \eqref{e:omegastar}, and $\ob^*$\label{p:obstar} is their arithmetic mean. We limit the domain of the integral \eqref{e:ntdbtrapu} to $K+\vt[\rb]>\mufs$, which has a negligible effect on results compared to the effect of increasing the chemical potential.

For the interacting gas, it is normal to consider the finite-size effect and mean-field interaction shift as independent additive corrections, which we do in \cite{Baillie09a}, but additional work is needed to find a consistent way of treating them together. We do not consider the finite-size effect due to factors other than the positive chemical potential.
\section{Numerical results\label{c:results}}\enlargethispage{1cm}
In this section we present results demonstrating the application of our mean-field theory to experimentally realistic regimes of a Bose gas in a 3D combined harmonic lattice potential. Our results quantify lattice and interaction effects on the thermal properties of the system. We refrain from discussing the critical temperature here, which we deal with in detail in \cite{Baillie09a}.
\subsection{Finite-size effect \label{s:fseffectresult}}
\begin{figure}[!t] 
  \setlength{\unitlength}{0.01\linewidth}
  \centering   
  \begin{picture}(80,80)
  \put(-10,0){\includegraphics{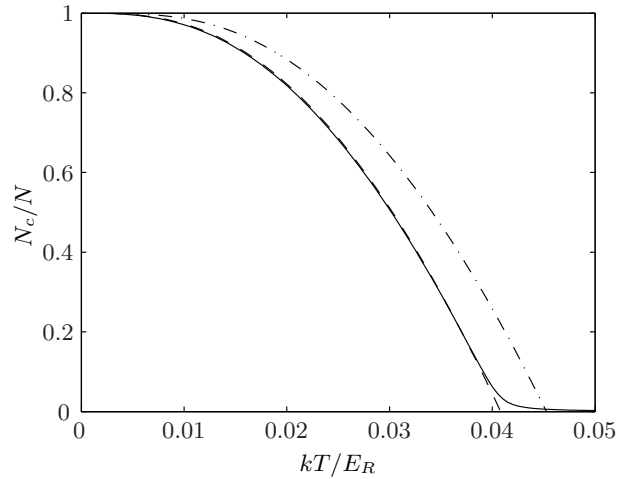} }
\end{picture}
  \caption{Condensate fraction for a non-interacting combined harmonic cubic-lattice in 3D with $N=1000$, $\omega = 0.02\omega_R$ and $V = 15E_R$, comparing full diagonalization (solid curve), LDA with $\mu \le \mufs$ (dashed curve) and LDA with $\mu\le 0$ (dashed-dotted curve).}
  \label{f:cfracfs}
\end{figure}
We consider the effect on the non-interacting condensate fraction of a non-zero ground-state energy. We plot the condensate fraction for $\omega = 0.02\omega_R$ and $V = 15E_R$ in Fig.~\ref{f:cfracfs} (results at other lattice depths and trap frequencies, are similar, except for scaling due to the different critical temperatures). We chose a small number of atoms, $N=1000$, to accentuate the finite-size effect. 

We see that the saturated chemical potential adjustment describes the bulk of the finite-size effect well, and the LDA calculation is in  excellent agreement with the full diagonalization (by diagonalization of $\Hl+\vt$ to obtain the ideal spectrum which is used solve for the condensate fraction using a grand-canonical approach, see \cite{Blakie07a}). We note that the LDA result shows a phase transition (i.e. discontinuous behavior) at the critical temperature, whereas the full diagonalization shows a more gradual change.
\subsection{Beyond nearest-neighbor hopping\label{s:bnnhresult}}
Here, we consider the effect on the non-interacting condensate fraction of beyond nearest-neighbor hopping (we use all neighbors for our numerical calculations in all other sections).

\begin{figure}[!t] 
  {\small\hspace{-7mm}\includegraphics{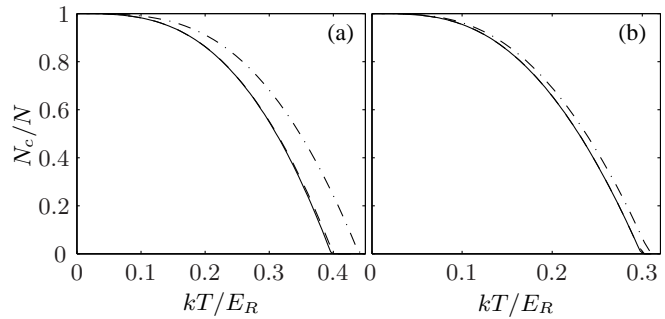}}
  \put(-130,105){(a)}
  \put(-20,105){(b)}
  \caption{Non-interacting condensate fraction for $N=10^5$, $\omega = 0.01\omega_R$, (a) $V=2E_R$ and (b) $V=5E_R$. The full diagonalization curve (solid curve) is almost obscured by the all-neighbor result (dashed curve) and is appreciably different from the nearest-neighbor result (dashed-dotted curve).}
  \label{f:cfracnn}
\end{figure}

We show the condensate fraction for $N=10^5$ and $\omega = 0.01\omega_R$ in Fig.~\ref{f:cfracnn}. We see that beyond nearest-neighbor hopping is significant for $V=2E_R$ and much less so for $V=5E_R$. For $V=10E_R$ (not shown), the condensate fractions are barely distinguishable on an equivalent plot. The decrease in significance of beyond nearest-neighbor hopping with increasing $V/E_R$, agrees with what we expect from Fig.~\ref{f:dostb3d} (see also appendix \ref{a:appwanhop}).

\subsection{Excited bands\label{s:exbandresult}}
In this section, we consider the significance of excited bands. We do not compare to the full diagonalization, since the separation into bands for that calculation is not well defined. The higher the temperature, the more important excited bands are, since they are more thermodynamically accessible. We therefore consider the significance of excited bands at the critical temperature. It is clear (e.g. see Fig.~\ref{f:dos15}) that increasing the lattice depth decreases the occupation for a given temperature, and hence the significance, of excited bands.

\begin{figure}[!t] 
  \centering   
  {\small\includegraphics{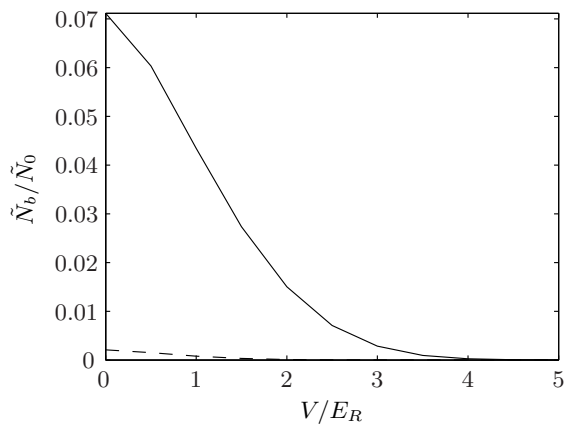}}
  \caption{Ratio of number of non-condensate atoms in first three (solid curve) and beyond first three (dashed curve) excited bands to non-condensate atoms in the ground band at the critical temperature for the experimental setup of \cite{Greiner02a}}
  \label{f:xbandfrac}
\end{figure}

We show the number of non-condensate atoms in excited bands as a proportion of the non-condensate number in the ground band in Fig.~\ref{f:xbandfrac}.
The calculations are for \Rb using HFBP with $a_s = 5.77\:\nano\metre$ and the parameters of \cite{Greiner02a} with an optical lattice wavelength of $\lambda = 2a=852\:\nano\metre$ and a spherical trap with frequency $\omega = 2\pi \times 24\:\hertz$. We used their maximum number of atoms, $N = 2\times 10^5$. We see that excited bands become insignificant for $V\gtrsim 3E_R$. The significance of excited bands at condensation would increase for an increased number of particles or a tighter trap, due to the increased critical temperature.
\subsection{Quantum depletion\label{s:qdeplresult}}
\begin{figure}[t] 
{\includegraphics[trim=110 120 25 70,clip=true,scale=0.75]{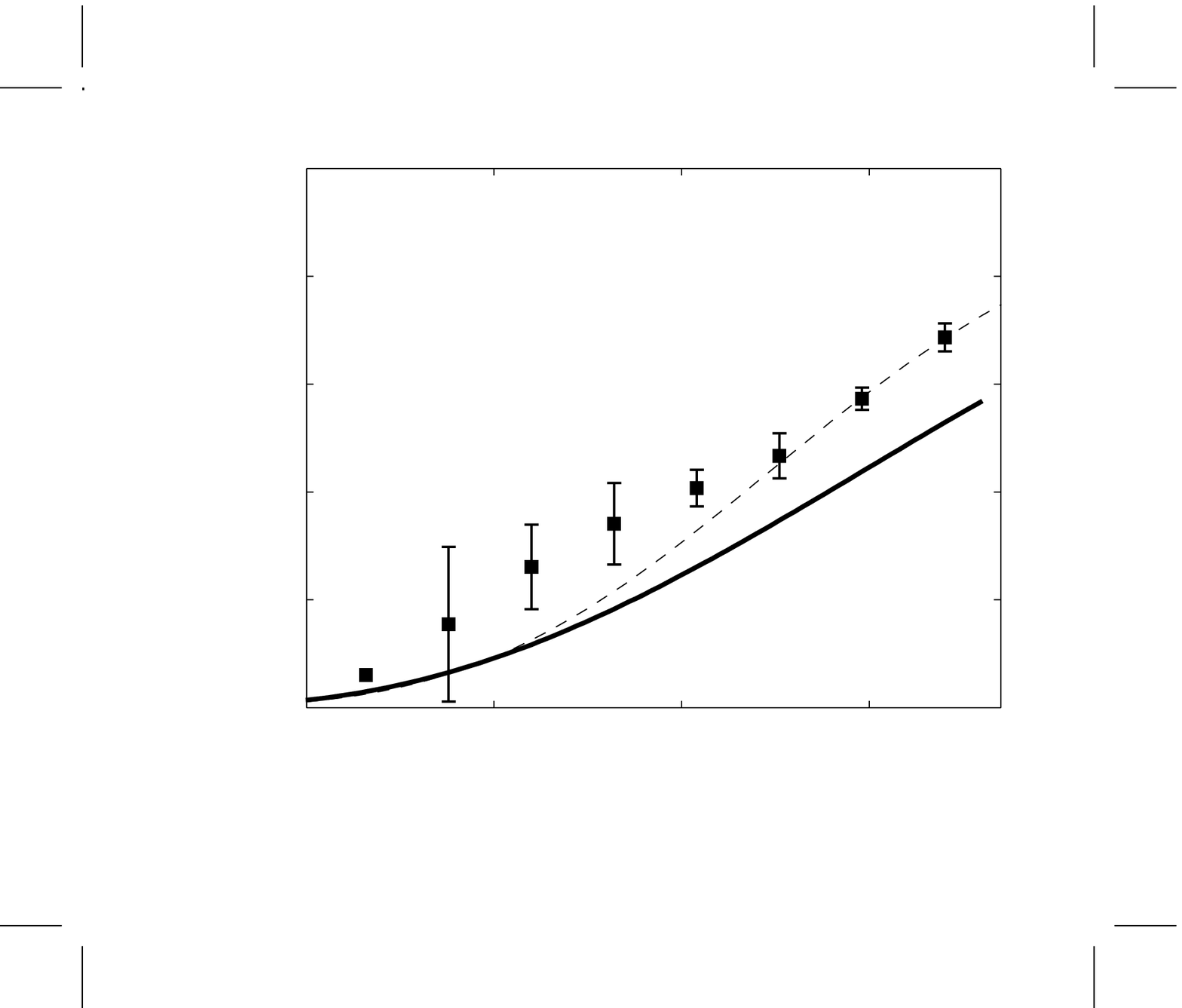}}
\put(-295,87){\rotatebox{90}{$\Nt/N$}}
\put(-155,-10){$V/E_R$}
\put(-277,8){0}
\put(-282,43){0.2}
\put(-282,78){0.4}
\put(-282,113){0.6}
\put(-282,148){0.8}
\put(-277,183){1}
\put(-270,0){5}
\put(-210,0){10}
\put(-148,0){15}
\put(-87,0){20}
\vspace{-2mm}\caption{Quantum depletion of \Na in a 3D optical lattice. The data points with error bars give the experimental quantum depletion. The curves give quantum depletion calculated by \cite{Xu06} (solid curve), our calculated quantum depletion (dashed curve).} \vspace{-3mm}
  \label{f:xuf2}
\end{figure}

The quantum depletion consists of the atoms promoted out of condensate due to interactions rather than thermal effects, thus leading to a reduction in the condensate fraction at $T=0$. The number of atoms in the quantum depletion is given by the temperature independent part of \eqref{e:nbdintk}:
\begin{align}
 N_{\mathrm{Q}} = \frac{1}{(2\pi)^d}\int\dbr\int_{\bz} \dbk  \,\norm{v_b(\bk,\br)}^2.
\end{align}
The quantum depletion is significantly enhanced by increasing the lattice depth which provides a convenient physical system to explore the crossover from a weakly to a strongly interacting Bose gas. The experimental measurement of quantum depletion in an optical lattice was reported in \cite{Xu06}. In that work, atoms were loaded into a lattice, which was linearly ramped up to a depth of $V\approx 20E_R$ and linearly ramped back down. By observing the diffuse background peak  of the momentum distribution of time-of-flight images during this sequence, the populations of the condensed and non-condensed atoms were estimated. The complete ramping procedure led to the production of $\sim 20\%$ thermal depletion (heating), and `Linear interpolation was used to subtract this small heating contribution (up to $10\%$ at the maximum lattice depth)' to obtain the quantum depletion \cite{Xu06}. Their results are presented in Fig.~\ref{f:xuf2}. 
We have calculated the zero temperature quantum depletion to compare with their experimental results. We have reproduced their calculations \cite{Xu06} with fixed peak density to a level indistinguishable on the plot (solid black curve), confirming our microscopic parameters agree with theirs, and we found that their results imply $N>10^7$ at $V=20E_R$. We used our LDA calculations with fixed total number\afn{We have assumed $N=1.7\times 10^5$ atoms, which is mentioned in \cite{Xu06}. Although the number of atoms throughout is unclear, using their maximum number of atoms, $N=5\times 10^5$, makes only a small change to the results.} rather than fixed peak density to give improved agreement with experimental results with no fitting parameters (dashed curve)\afn[.]{We note that our methods are not valid after the Mott-insulator transition. Although the $n=1$ Mott-insulator transition is at $V=16.4E_R$, the `measurements were performed at a peak lattice site occupancy number $\sim 7$' \cite{Xu06}, and the Mott-insulator transition is at $V>20E_R$ for $n\ge 3$, which extends our validity regime somewhat.} The agreement is improved over the entire range, most noticeably at higher lattice depths. More precise experimental measurements at intermediate lattice depths to better test theory would be useful.\vspace{-6mm}
\subsection{Effect of quasi-particles}
\begin{figure}[t] 
  \centering   
	{\small \includegraphics{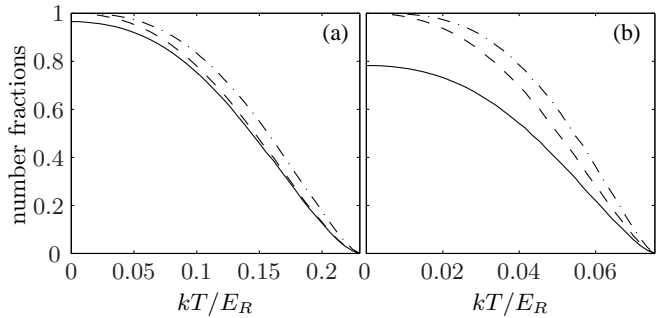}}
  \put(-130,105){(a)}
  \put(-20,105){(b)}\vspace{-4mm}
\caption{Condensate and quantum depletion fractions for the parameters of \cite{Greiner02a}, (a) $V=5E_R$ and (b) $V=10E_R$. Results are for the HFBP method for the condensate only (solid curve), for the condensate plus quantum depletion (dashed curve) and for the Hartree-Fock method (dashed-dotted curve).}
 \label{f:qpartcf}
\end{figure}

\begin{figure}[t] 
  \centering   
	{\small \includegraphics{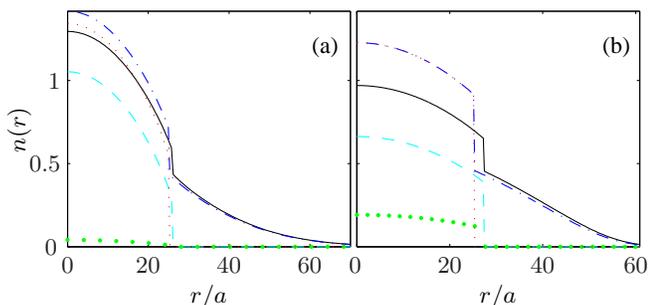}}
	  \put(-130,97){(a)}
  \put(-20,97){(b)}\vspace{-4mm}
\caption{(Color online) Spatial densities for the parameters of \cite{Greiner02a} at $T=0.8T_c$, (a) $V=5E_R$ and (b) $V=10E_R$. Results are for the HFBP method: total (black solid curve), condensate (cyan dashed curve), quantum depletion (green filled circles); and the Hartree-Fock method: total (blue dashed-dotted curve) and condensate (red dotted curve).}
 \label{f:qpartn}
\end{figure}

In addition to the quantum depletion, which was considered at zero temperature in section \ref{s:qdeplresult}, the Bogoliubov quasi-particles modify the energy dispersion as in \eqref{e:sce}. We compare the quantum depletion to the residual Bogoliubov effect in this section (using the parameters of \cite{Greiner02a}, as discussed in section \ref{s:exbandresult}). In Fig.~\ref{f:qpartcf}, we show the condensate fraction and the condensate plus quantum depletion fraction. At zero temperature, the only effect of quasi-particles is the quantum depletion. The methods with and without quasi-particles give the same results above the critical temperature and the same critical temperature\afn[,]{The critical temperature is the same if we define it as the lowest temperature for which all particles can be accommodated as thermal atoms. We note the consistency issues near the critical temperature discussed in \cite{Gies04}.} since equations \eqref{e:Ebtrapu} and \eqref{e:ntdbtrapu} are the same when there is no condensate. In Fig.~\ref{f:qpartcf} we can see the zero temperature increase in quantum depletion due to the increase in lattice depth (as in Fig.~\ref{f:xuf2}) and we can see that the nature of the Bogoliubov quasi-particle spectrum \eqref{e:sce} also increases thermal depletion relative to the Hartree-Fock prediction. 
In Fig.~\ref{f:qpartn} we show the total spatial density, and that of the condensate and quantum depletion. 
The quantum depletion follows the condensate density from \eqref{e:sce} and \eqref{e:vsqkr}. A larger lattice depth increases the effective interaction, decreasing the core density and, for the Hartree-Fock case, forces all of the thermal depletion away from the condensate region.
\section{Conclusions}
The main purpose of this paper has been the derivation of an accurate, computationally tractable theory for describing experiments with finite temperature Bose gases in optical lattices. Based on an extended Bose-Hubbard model, derived from the full cold atom Hamiltonian, our theory includes the important physical effects needed to describe this system over a wide parameter regime. We obtain a mean-field theory for the system using the Hartree-Fock-Bogoliubov-Popov approximation. Through the development of two key techniques, a local density approximation for the lattice physics and an envelope approximation for spatial dependence of the mean fields, we realize a formalism for calculation that is efficient and accurate. By neglecting the extended features our formalism we show that it reduces to a form equivalent to the Bose-Hubbard mean-field theory of \cite{Lin08}.

We have presented a range of results verifying the accuracy of our theory, and demonstrating the regimes in which extended features of our model, over the usual Bose Hubbard model, are important. We have also compared to recent experimental results by the MIT group, and find that our formalism provides improved agreement with the experimental data over previous calculations \cite{Xu06}.

The methods outlined in this paper can be applied to other thermodynamic quantities. For example, we have used our  numerical results to calculate the entropy:
\begin{align}
\frac{S}{\kb} &= \sum_b \int\dbr \int \diff K g_b(K) 
\left\{ \beta E_b(K,\br)\nbe\left[E_b(K,\br)\right] \bsplit \hspace{3cm} - \ln\left[ 1-e^{-\beta E_b(K,\br)} \right] 
\right\},
\end{align}
and from that the specific heat and then the energy, can be obtained. Our formulation is amenable to analytical results as we have done in \cite{Baillie09a}.

Experimental work in optical lattices is continuing apace and, with the recent development of thermometry techniques \cite{McKay09}, it is likely that thermodynamics will be measured in the near future. For the purposes of developing better understanding of lattice bosons, and the emergence of beyond mean-field effects, it is crucial to have a quantitative and accurate mean-field theory for comparison. The theory presented here serves this purpose.
\begin{acknowledgments}
The authors acknowledge support from  the University of Otago Research Committee and NZ-FRST contract NERF-UOOX0703, and useful discussions with Ashton Bradley.
\end{acknowledgments}
\appendix
\section{Wannier functions\label{a:appwannier}}
\begin{figure}[t] 
  \centering   
  {\small\includegraphics{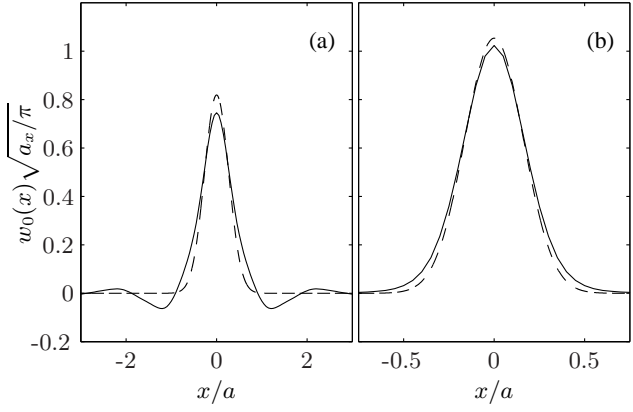}}
    \put(-125,135){(a)}
  \put(-20,135){(b)}
  \caption{Ground-band Wannier functions (solid curve) compared to the Gaussian approximation (dashed curve) for (a) $V=2E_R$ and (b) $V=15E_R$}
  \label{f:wannierg0}
\end{figure}

\begin{figure}[t] 
  \centering   
  {\small\includegraphics{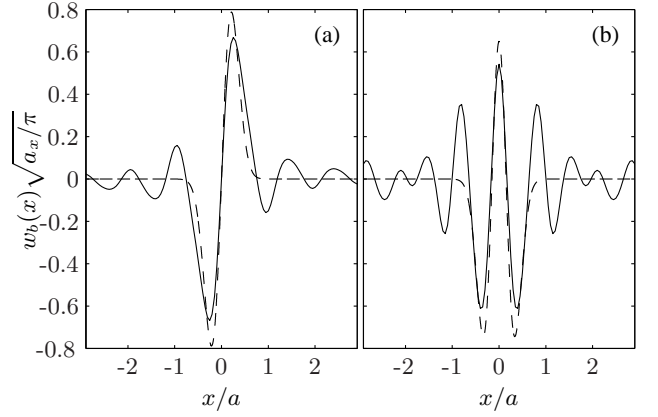}}
    \put(-125,140){(a)}
  \put(-20,140){(b)}
  \caption{Wannier function for the (a) first and (b) second excited bands for $V=5E_R$ (solid curve) compared to the harmonic oscillator approximation (dashed curve)}
  \label{f:wannierex}
\end{figure}

We define the Wannier function for band $b$, localized at site $\bR_i$ as:
\begin{equation}
     w_b(\br-\bR_i) \equiv \frac{1}{\sqrt{\ns}} \sum_{\bk\in \bz} e ^ {-\ii \bk \cdot \bR_i} \psi_{b,\bk}(\br), \label{e:wandef}
\end{equation}
where $\ns$ is the number of sites (we let $\ns\rightarrow\infty$ for the combined harmonic lattice). We have:
\begin{align}
   \psi_{b,\bk}(\br) &= \frac{1}{\sqrt{\ns}}\sum_{i=1}^{\ns}  e^{\ii \bk \cdot \bR_i}   w_b(\br-\bR_i). \label{e:psisumw}
\end{align}
For $\bR_i$ on the lattice, $
%
    \sum_{\bk\in \bz} e^{\ii \bk \cdot \bR_i} 
  = \ns\delta_{\bR_i,\bzero}$, 
%
so we have:
\begin{align}
  \int \dbr \, w^*_b(\br-\bR_i) w_{b'}(\br-\bR_{i'}) 
  &= \delta_{bb'} \delta_{ii'}.  \label{e:onormw}
\end{align}
For an optical lattice in 1D, we show the Wannier function for the ground band in Fig.~\ref{f:wannierg0} and for the first and second excited bands in Fig.~\ref{f:wannierex}. The harmonic oscillator approximation (the eigenstates of $\vl \approx \sum_{j=1}^3 V_j (\pi r_j/a_j)^2$) overstates the peak height at the expense of the tails, and misses the detailed structure of the Wannier functions. 
\section{Hopping matrix\label{a:appwanhop}}
\begin{figure}[t] 
  \centering   
  {\small\includegraphics{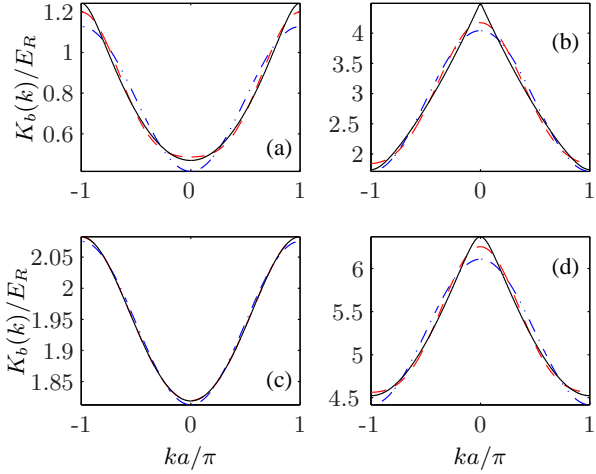}}
  \put(-133,120){(a)}
  \put(-25,160){(b)}
  \put(-133,32){(c)}
  \put(-25,75){(d)}
  \caption{(Color online) Fourier series for the 1D \unif lattice spectrum for (a,b) $V=E_R$, [(c),(d)]  $V=5E_R$, [(a),(c)] ground band and [(b),(d)] first excited band, using all neighbors (black solid curve),   
  nearest and next nearest neighbors (red dashed curve) and nearest neighbors (blue dashed-dotted curve)}
  \label{f:dispcos}
\end{figure}

Since $\Hl\psi_{b,\bk}(\br) = K_b(\bk)\psi_{b,\bk}(\br)$, we have (as in \cite{Ziman76}) 
%
 $  \Hl w_b(\br-\bR_{i'})
    =  -\sum_{i'=1}^{\ns} J_{b,i,i'} w_b(\br-\bR_{i})$, 
%
where hopping matrix, defined as \eqref{e:jdef}:
\begin{align}
   J_{b,i,i'}
   = -\frac{1}{\ns}
    \sum_{\bk\in {\bz}} e ^ {-\ii \bk \cdot (\bR_{i'}-\bR_{i})} K_b(\bk), \label{e:jni}
\end{align}
is the Fourier transform of the energy. In particular, $-J_{b,i,i} =  \sum_{\bk\in {\bz}}K_b(\bk)/\ns$ is the average energy in the band. So,
\begin{align}
   &\int\dbr\, w^*_{b}(\br-\bR_i) \Hl w_{b'}(\br-\bR_{i'}) \notag\\
   &=  \frac{\delta_{bb'}}{\ns} \sum_{\bk\in {\bz}} e ^ {-\ii \bk \cdot \left(\bR_{i'}-  \bR_{i}\right)} K_{b}(\bk) ,  \label{e:Jn1n2}
\end{align}
so that there is no inter-band hopping and the hopping matrix depends only on the difference $\bR_i-  \bR_{i'}$. We can invert \eqref{e:jni} to write the dispersion relation as a Fourier series:
\begin{align}
  K_{b}(\bk) = -\sum_{i'=1}^{\ns} J_{b,i,i'} \,e^{\ii \bk \cdot (\bR_{i'}-\bR_{i})} = -\sum_{i=1}^{\ns} J_{b,i,0} \,e^{-\ii \bk \cdot \bR_i}. \label{e:enkF}
\end{align}
For the 1D case, if the spectrum is even in $k_x$ then:
\begin{align}
   K_{b_x}(k_x) &= -J^0_{b_x,x}-2\sum_{l>0} J^l_{b_x,x} \,\cos(l k_x a_x).
\end{align}
We demonstrate the Fourier cosine series for the \unif lattice spectrum in Fig.~\ref{f:dispcos}. For $V=E_R$, we can see that a few terms are needed for the series to approach the nearly free-particle dispersion. By $V=5E_R$, the ground band is well described by nearest neighbors. For the first excited band, the approach to nearest-neighbor dispersion with increasing $V/E_R$ is somewhat slower.
The width of band $b_x$ is:
\begin{align}
   \norm{K_{b_x}\fracb{\pi}{a_x} - K_{b_x}(0)} 
    = 4\norm{\sum_{l>0} J^{2l-1}_{b_x,x}},\label{e:bandwidth}
\end{align}
so, for a separable lattice:
\begin{align}
K_b(\bk) &= -\sum_{j=1}^d \left[ J^0_{b_j,j} + 2\sum_{l>0} J^l_{b_j,j} \cos(l k_j a_j) \right],\label{e:enkfouriercos}
\end{align}
and the width of band $b$ is:
\begin{align}
K_b^{\max} - \Kbm &= 4\sum_j\norm{\sum_{l>0}J^{2l-1}_{b_j,j}}.\label{e:widthj}
\end{align}
In the tight-binding case where $l=1$ dominates, the bandwidth is 
$4\sum_j\norm{ J^1_{b_j,j}}$.

\begin{figure}[t]
  \centering   
  {\small\includegraphics{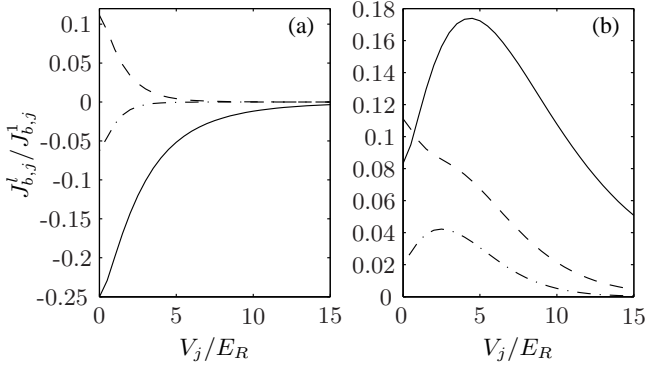}}
  \put(-142,125){(a)}
  \put(-27,125){(b)}
  \caption{Ratio of beyond nearest-neighbor to nearest-neighbor hopping. $J^2_{b,j}/J^1_{b,j}$ (solid curve) , $J^3_{b,j}/J^1_{b,j}$ (dashed curve), $J^4_{b,j}/J^1_{b,j}$ (dashed-dotted curve) for (a) ground band and (b) first excited band}
  \label{f:jdj}
\end{figure}

The ratio of beyond nearest-neighbor to nearest-neighbor hopping in shown in Fig.~\ref{f:jdj} and we see that the ground-band next-nearest-neighbor hopping matrix element is as much as $25\%$ of its nearest-neighbor counterpart at $V_j=0$, but decreases rapidly with increasing $V_j$. Beyond next-nearest-neighbor hopping is less significant. For the first excited band, some of the ratios can increase initially.

\section{Harmonic trap\label{a:trap}}
In this work, we will always use the local energy form \eqref{e:videf} to represent the harmonic trap. In this section, we consider an exact treatment for the separable case, by defining:
\begin{align}
  v_{b,b',i,i'} \equiv \int\dbr\, \vt w^*_{b}(\br-\bR_i) w_{b'}(\br-\bR_{i'}). 
\end{align}
\subsection{On-site variation}
Here we consider the accuracy of \eqref{e:videf} to the diagonal part of $v_{b,b',i,i'}$. There are three components to the integral in \eqref{e:videf}, one for each trap direction and the three components are additive. Considering, e.g., the $x$ component, we have (using $X_i$ for the $x$ component of $\bR_i$):
\begin{align}
   &   \int\dbr\, \half m \omega_x^2 x^2 \norm{w_{b}(\br-\bR_i)}^2\notag\\
 &    = \half m \omega_x^2 X_i^2  + \half m \omega_x^2\intinf\diff x\, x^2\norm{w_{b}(x)}^2      ,
\end{align}
since $x\norm{w_{b}(x)}^2$ is odd and $w_b(\br)$ is normalized. For the ground band, we can recover \eqref{e:videf} by absorbing a constant into the chemical potential. For excited bands there is an error due to the difference $ \half m \omega_x^2\intinf\diff x\, x^2\left[\norm{w_{b}(x)}^2- \norm{w_{0}(x)}^2\right]$, 
which is applied to $\nh_{b,i}$ in the Hamiltonian. We plot the contribution for the first excited band in Fig. \ref{f:trapall}(a).
\subsection{Off-site contribution}
\begin{figure}[!t] 
  {\includegraphics[scale=0.98]{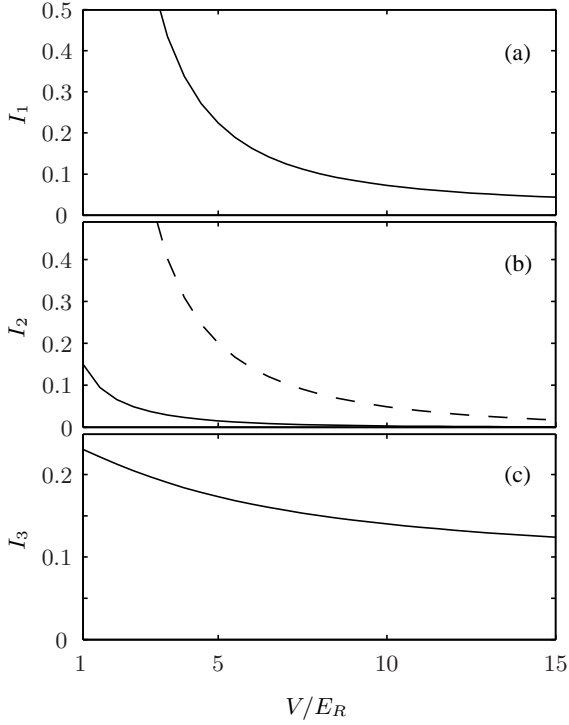}
  \put(-35,250){(a)}
  \put(-35,170){(b)}
  \put(-35,90){(c)}}
  \caption{Error due to (a) on-site variation of the trap,  $I_1 = \intinf\diff x\, (x/a_x)^2\left[\norm{w_{1}(x)}^2- \norm{w_{0}(x)}^2\right]$ (this form is chosen so that the error is in units of $\half m \omega_x^2 a_x^2 $), 
  (b) contribution from adjacent sites, $I_2 =  \norm{\intinf\diff x\, (x/a_x)^2 w^*_{b}(x) w_{b}(x-a_x)}$, ground band (solid curve), first excited band (dashed curve), (c) Wannier function overlap between the ground and first excited bands, 
  $I_3 = \intinf\diff x\, (x/a_x) w^*_{0}(x) w_{1}(x)$. 
  }
  \label{f:trapall}
\end{figure}
Now, we consider the case with $i\ne i'$ and $b=b'$. We note again that the components of the trap contributing to the integral in the three directions are additive. We only get a potential error in the $x$ component if components in the other directions of $i$ and $i'$ are equal. Then, for $X_i\ne X_{i'}$:
\begin{align}
    &    \int\dbr\, \half m \omega_x^2 x^2 w^*_{b}(\br-\bR_i) w_{b}(\br-\bR_{i'})\notag\\
  &          = \half m \omega_x^2 \intinf\diff x\, x^2 w^*_{b}(x) w_{b}(x-(X_{i'}-X_i)),
\end{align}
since $w_{b}(x-X_i)$ and $w_{b}(x-X_{i'})$ are orthogonal and $w^*_{b}(x-X_i) w_{b}(x-X_{i'})$ is even about $(X_i+X_{i'})/2$ as $w_b(x)$ is either even or 
odd. 
In Fig.~\ref{f:trapall}(b) we plot this contribution for nearest neighbors as a function of $V$. 
\subsection{Inter-band contribution}
Now we consider the case with $b\ne b'$ and $i=i'$. To allow for this contribution, it would be necessary to include matrix elements between bands in the Hamiltonian.

To quantify the error, we consider the additive component in the $x$ direction. There is only a contribution if the other components of band $b$ and $b'$ are equal. Then, with $b_x$ being the $x$ component of $b$ and $b_x\ne b'_x$:
\begin{align}
  &        \int\dbr\,  \half m \omega_x^2 x^2 w^*_{b}(\br-\bR_i) w_{b'}(\br-\bR_{i}) \notag\\
        &= \half m \omega_x^2 \intinf\diff x\,  (x+X_i)^2 w^*_{b_x}(x) w_{b'_x}(x) .
\end{align}
Considering, e.g. $b_x=0$ (the ground band), and $b'_x=1$ (the first excited band) $w^*_{0}(x) w_{1}(x)$ is odd so the above becomes $m \omega_x^2 X_i \intinf\diff x\, x w^*_{0}(x) w_{1}(x)$. In Fig.~\ref{f:trapall}(c), we plot this contribution as a function of $V$.
\section{Interaction coefficients\label{a:intcoeff}}
\subsection{Beyond the on-site interaction approximation \label{a:offsite}}
Here we derive approximate results for interactions extending to all sites. To do this, we make the HFBP mean-field approximations, as discussed in section \ref{c:mfield}, but starting from the more general extended Bose-Hubbard Hamiltonian \eqref{e:Kh}. As in the on-site case, we ignore collisional couplings between bands in the many body-state. For the non-condensate, we also ignore collisional coupling that relies on coherences between sites (i.e. requiring two indices at two sites) in the many-body state, to find:
\begin{align}
  &\sum_{\myatop{i_1,i_2,i_3,i_4}{b_1,b_2,b_3,b_4}} \dehd_{b_1,i_1}\dehd_{b_2,i_2}\deh_{b_3,i_3}\deh_{b_4,i_4} 
  U_{\myatop{i_1,i_2,i_3,i_4}{b_1,b_2,b_3,b_4}} 
  \notag\\&\approx 4\sum_{i,b,b'} 
  \dehd_{b,i}\deh_{b,i}\sum_{i'}\nt_{b',i'} U_{\myatop{i,i',i,i'}{b,b',b,b'}}.\label{e:sumbdeh}
\end{align}
We assume that the density varies sufficiently slowly that $\nt_{b,i} \approx \nt_{b,j}$ for sites $\bR_j$ near $\bR_i$. In the following, we will sum over all sites, by assuming that where the approximation $\nt_{b,i} \approx \nt_{b,j}$ is poor, due to the sites being far apart, these terms will be suppressed by the negligible Wannier function overlap. Then we have:
\begin{align}
 &\sum_{\myatop{i_1,i_2,i_3,i_4}{b_1,b_2,b_3,b_4}} \dehd_{b_1,i_1}\dehd_{b_2,i_2}\deh_{b_3,i_3}\deh_{b_4,i_4} U_{\myatop{i_1,i_2,i_3,i_4}{b_1,b_2,b_3,b_4}} \notag\\
  &\approx 4g\sum_{i,b,b'} \dehd_{b,i}\deh_{b,i}\nt_{b',i}
  \sum_{i'}\int\dbr \norm{w_b(\br) w_{b'}(\br-\bR_{i'})}^2 \notag\\
  &= 4\sum_{i,b,b'} \dehd_{b,i}\deh_{b,i}\nt_{b',i}U'_{bb'},
\end{align}
which is the same as in \eqref{e:K2bidef} with $U'_{bb'}$ substituted for $U_{bb'}$ where:
\begin{align}
U'_{bb'} \equiv  g\sum_{i'}\int\dbr \norm{w_b(\br) w_{b'}(\br-\bR_{i'})}^2.\label{e:Uddef}
\end{align}
For the coherent condensate, we assume that $z_i \approx z_j$, for sites $\bR_j$ near $\bR_i$. As above, we assume that contributions between sites far apart are suppressed by the negligible Wannier function overlap. Assuming that the phase factors are chosen so that $w_0(\br)$ is real, 
we have, for site $\bR_{i_1}$:
%
\begin{align}
 &\sum_{i_2,i_3,i_4} z_{i_1}^* z_{i_2}^* z_{i_3} z_{i_4} U_{\myatop{i_1,i_2,i_3,i_4}{0,0,0,0}}\notag\\
 &= g\sum_{i_2,i_3,i_4} z_{i_1}^* z_{i_2}^* z_{i_3} z_{i_4} \int\dbr\,
 \prod_{j=1}^4 w_0(\br-\bR_{i_j}) \notag\\
  &\approx 
  g\norm{z_{i_1}}^4 \hspace{-2mm}\sum_{i_2,i_3,i_4} \int\dbr\, 
 \prod_{j=1}^4 w_0(\br-\bR_{i_j}) \notag\\
  &=   g\norm{z_{i_1}}^4 \int\dbr\,   w_0(\br)\left[ \sqrt{\ns}\psi_{0,\bzero}(\br)\right]^3
  =   \norm{z_{i_1}}^4U''_{00},\label{e:Uwaninterf}
\end{align}
%
where $\sum_{i} w_b(\br-\bR_i) = \sqrt{\ns}\psi_{b,0}(\br)$ is the Bloch function normalized over a single site, from \eqref{e:psisumw}, and $\psi_{0,\bzero}(r_i) \propto \ce_0(r_i\pi/a_i,q)$ (the Mathieu function) is real and periodic on the lattice. The result takes the same form as above with $U''_{00}$ substituted for $U_{00}$ where:
\begin{align}
U''_{00} \equiv g\int\dbr\,   w_0(\br)\left[\sqrt{\ns} \psi_{0,\bzero}(\br)\right]^3.\label{e:Udd}
\end{align}
Similar arguments could be used for the terms involving interactions between the condensate and the non-condensate. The above results are appropriate for the pure thermal gas, e.g. for finding the critical temperature from above, and for the pure condensate at zero temperature. To quantify the effect of off-site interactions on the thermal depletion, terms for interactions between the condensate and the non-condensate would be needed.
\subsection{No lattice limit\label{s:intnll}}
When there is no lattice, the Hamiltonian \eqref{e:Hnilatt} gives us $K_b(\bk) = \hbar^2k^2/2m$ and the Bloch states are plane waves. Using these to evaluate the Wannier functions from \eqref{e:wandef}, and then the all-sites interaction coefficients: $U''_{00}$ easily from \eqref{e:Udd}, and $U'_{bb'}$ from \eqref{e:Uddef} by splitting the sum into axial components and recognizing the Riemann zeta sums to get:
\begin{align}
  U''_{00} &= U'_{00} = U'_{000,001} = U'_{001,001} = U'_{010,001} = \frac{g}{a^d}.\label{e:Unolattallsites}
\end{align}
\begin{figure}[!t] 
  \centering   
  {\small\includegraphics{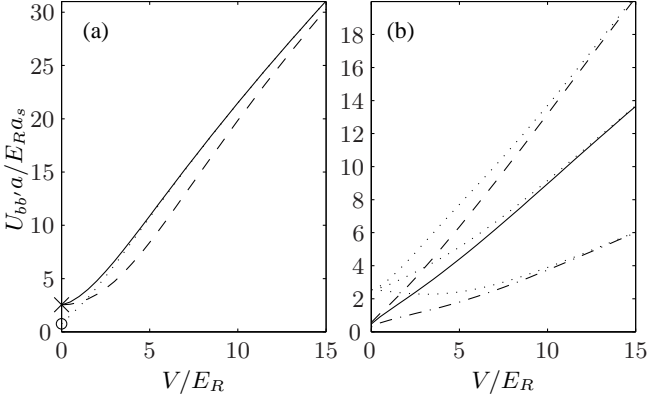}
    \put(-215,135){(a)}
    \put(-100,135){(b)} }
  \caption{Interaction coefficients in 3D.  (a) Ground band. 
On site (dotted curve). All sites: non-condensate $U'$ \eqref{e:Uddef} (solid curve), condensate $U''$ \eqref{e:Udd} (dashed curve). No lattice limit: all sites \eqref{e:Unolattallsites} ($\times$), on site \eqref{e:Unolatt} ($\circ$). (b) Excited-band. On-site: 000,001 (these integers specify the components $b_x,b_y,b_z$ of each band) (solid), 001,001 (dashed curve), 010,001 (dashed-dotted curve); corresponding all sites (dotted curve).}
  \label{f:Ugrndex}
\end{figure}
So that, if we use all-site interaction coefficients and also treat $\ncd(\br)$ and $\ntd{b}(\br)$ as the condensate and non-condensate densities (rather than as envelope functions, with densities defined by \eqref{e:ncdef} and \eqref{e:ntpsidef}, 
although the total condensate and non-condensate numbers do not depend on this distinction, from \eqref{e:Ncintcalc} and \eqref{e:Ntintcalc}) then all of our LDA equations in section \ref{c:lda} would be the same as we would get from a no lattice calculation \cite{Giorgini97b}, in spite of our expansion of the field operators in a Wannier basis. When only on-site interactions are included there is a shortfall, using \eqref{e:Ubb}:
\begin{gather}
   U = \frac{g}{a^d}\fracb{2}{3}^d\csp
   U_{000,00n} = \frac{g}{a^3}\frac{5}{27}\csp\notag\\
   U_{00n,00n} = \frac{g}{a^3}\frac{2}{9}\csp
   U_{0n0,00n} = \frac{g}{a^3}\frac{25}{216},
    \label{e:Unolatt}
\end{gather}
of, for example, $1-(2/3)^3 = 70\%$ for the 3D ground-band coefficient. For reference in Fig.~\ref{f:Ugrndex}, $a/E_Ra_s = 8a^3/g\pi$.\vspace{0.5cm}

\subsection{Comparison}
The 3D ground-band interaction coefficients are shown in Fig.~\ref{f:Ugrndex}(a). Both all-sites interaction coefficients, $U'_{00}$ and $U''_{00}$, include their corresponding on-site component, $U_{00}$, in their sums, \eqref{e:Uddef} and \eqref{e:Uwaninterf}. For the non-condensate interaction coefficient, all other terms in the sum are positive (since we have excluded interference), so that off-site interactions always increase the interaction coefficient (relative to $U$).

The 3D excited-band interaction coefficients are shown in Fig.~\ref{f:Ugrndex}(b). The results all tend to the expected limits at $V=0$. The gap between all-site and on-site interaction coefficients is maintained for higher $V/E_R$ than for the ground-band, since the excited-band Wannier functions are less localized.
\begin{widetext}

\section{Diagonalization of the quadratic Hamiltonian\label{a:epaps}}

This appendix gives a derivation of the quadratic Hamiltonian \eqref{e:Kq}, and a proof that the Bogoliubov-de Gennes equations reduce the quadratic Hamiltonian to diagonal form \eqref{e:Khdiag}. 

\subsection{Quadratic Hamiltonian}
We begin with the extended Bose-Hubbard Hamiltonian:
\begin{align}
  \Kh_i &\equiv \sum_{b}\left[-\sum_{i'}\left(J_{b,i,i'}\ahd_{b,i}\ah_{b,i'} \right) + \nh_{b,i}(\vi-\mu)\right] 
  + \frac{1}{2}\sum_{{b_1,b_2,b_3,b_4}} \ahd_{b_1,i}\ahd_{b_2,i}\ah_{b_3,i}\ah_{b_4,i}U_{\myatop{i,i,i,i}{b_1,b_2,b_3,b_4}}. \label{e:EPAPSKhi}
\end{align}
We make the substitutions $ z_i \equiv \av{\ah_{0,i}}\csp\deh_{0,i}\equiv \ah_{0,i}-z_i$ 
for the ground band, and $\deh_{b,i}\equiv\ah_{b,i}$ above the ground band (with the operators $\deh_{b,i}$ satisfying standard bosonic commutation relations) into the interaction term of \eqref{e:EPAPSKhi} to obtain:
\begin{align}
&\half   \sum_{b_1,b_2,b_3,b_4}\ahd_{b_1,i}\ahd_{b_2,i}\ah_{b_3,i}\ah_{b_4,i}\,U_{\myatop{i,i,i,i}{b_1,b_2,b_3,b_4}}
   = \half\norm{z_i}^4U_{00} + \sum_{b}  \left( z_i^* \deh_{b,i} + z_i \dehd_{b,i}\right)\norm{z_i}^2U_{\myatop{i,i,i,i}{0,0,0,b}}\notag\\
   &+\sum_{b,b'}\left(\half z_i^{*2}\deh_{b,i}\deh_{b',i} + \half z^2_i\dehd_{b,i}\dehd_{b',i} + 2\norm{z_{i}}^{2}\dehd_{b,i}\deh_{b',i}\right)U_{\myatop{i,i,i,i}{0,0,b,b'}}
   +\sum_{b_1,b_2,b_3}\left(z_i^* \dehd_{b_1,i}\deh_{b_2,i}\deh_{b_3,i} + z_i \dehd_{b_1,i}\dehd_{b_2,i}\deh_{b_3,i}\right)U_{\myatop{i,i,i,i}{0,b_1,b_2,b_3}}\notag\\
 &+ \half\sum_{b_1,b_2,b_3,b_4} \dehd_{b_1,i}\dehd_{b_2,i}\deh_{b_3,i}\deh_{b_4,i}U_{\myatop{i,i,i,i}{b_1,b_2,b_3,b_4}},
\end{align}
where we have assumed phase factors are chosen so that the Wannier functions are real, so that the order of subscripts in $U_{\myatop{i,i,i,i}{b_1,b_2,b_3,b_4}}$ is unimportant.

We make a quadratic Hamiltonian simplification by making a mean-field approximation motivated by Wick's theorem \cite{Griffin96}. For the fourth order terms, we find:
\begin{align}
  \half\sum_{b_1,b_2,b_3,b_4} \dehd_{b_1,i}\dehd_{b_2,i}\deh_{b_3,i}\deh_{b_4,i}U_{\myatop{i,i,i,i}{b_1,b_2,b_3,b_4}} \approx 2\sum_{b,b'}\nt_{b',i} \dehd_{b,i}\deh_{b,i}U_{bb'},
\end{align}
where $  U_{bb'} \equiv g\int\dbr\, \norm{w_{b}(\br)w_{b'}(\br)}^2$ and we have used a Popov approximation to eliminate the terms $\av{\dehd_{b, i} \dehd_{b, i}}$ and $\av{\deh_{b, i} \deh_{b, i}}$, and we neglect pairs with different band indices, since 
we ignore collisional couplings between bands in the many-body state. Similarly, we simplify the third order terms by analogy with Wick's theorem \cite{Morgan00} to find:
\begin{align}
  \sum_{b_1,b_2,b_3}z_i^*\dehd_{b_1, i}\deh_{b_2, i}\deh_{b_3, i}U_{\myatop{i,i,i,i}{0,b_1,b_2,b_3}} \approx 2z^*_{i}\deh_{0, i}\sum_b\nt_{b, i}U_{0b},
\end{align}
and the adjoint of this equation. We set
the linear terms
$\left( z_i^* \deh_{b,i} + z_i \dehd_{b,i}\right)\norm{z_i}^2$ to zero for $b\ne0$ and 
the quadratic terms 
$\norm{z_{i}}^{2}\dehd_{b,i}\deh_{b',i}$,
$z_i^{*2}\deh_{b,i} \deh_{b',i}$ and
$z^2_i  \dehd_{b,i}\dehd_{b',i}$ 
to zero for $b\ne b'$ by the same assumption that interactions are perturbative relative to the band-gap energy scale. Our interaction term becomes:
{
\begin{align}
 \half \sum_{b_1,b_2,b_3,b_4} &\ahd_{b_1,i}\ahd_{b_2,i}\ah_{b_3,i}\ah_{b_4,i}\,U_{\myatop{i,i,i,i}{b_1,b_2,b_3,b_4}}
  \approx \left(\half\norm{z_{i}}^2 + z_i^*\deh_{0,i} + z_i\dehd_{0,i}\right) \norm{z_i}^2 U_{00}\notag\\ 
 &+ \sum_b\left(\half z_{i}^{*2}\dehs_{b,i} + \half z^2_{i}\dehds_{b,i} + 2\norm{z_{i}}^{2}\dehd_{b,i}\deh_{b,i}
 + 2z^*_{i} \nt_{b,i}\deh_{0, i}
 + 2z_i  \nt_{b,i}\dehd_{0, i} \right)U_{0b}
 + 2\sum_{b,b'} \nt_{b',i}\dehd_{b,i}\deh_{b,i} U_{bb'},\label{e:EPAPSintquad}
\end{align}
}
which gives:
\begin{align}
  \Kq \equiv \sum_i \left(\Kh_{0,i} + \Kh_{1,i} + \Khd_{1,i} + \sum_b \Kh_{2,b,i}\right),\label{e:EPAPSKq}
\end{align}
with:
\begin{align}
  \Kh_{0,i} &\equiv z_i^*\left(-\sum_{i'} J_{0,i,i'} \Sh{i',i}  + \vi - \mu +  \frac{U_{00}}{2}\norm{z_i}^2\right)z_i,\\
  \Kh_{1,i} &\equiv \dehd_{0,i}\left(-\sum_{i'} J_{0,i,i'}\Sh{i',i} + \vi - \mu +
   U_{00}\norm{z_i}^2 + 2\sum_{b} U_{0b}\nt_{b,i} \right)z_i,\\
  \Kh_{2,b,i} &\equiv \dehd_{b,i}\Lh_{b,i}\deh_{b,i} + \frac{U_{0b}}{2} \left(\dehds_{b,i}z^2_{i} + \dehs_{b,i}z_{i}^{*2}  \right),\label{e:EPAPSK2bidef}
\end{align}
where:
\begin{align}
  \Lh_{b,i} &\equiv -\sum_{i'} J_{b,i,i'}\Sh{i',i} + \vi - \mu 
  + 2U_{0b}\norm{z_i}^2 + 2\sum_{b'}U_{bb'}\nt_{b',i},\label{e:EPAPSLdef}
\end{align}
and $\Sh{i',i}$ is the shift operator from the site $\bR_i$ to $\bR_{i'}$, e.g. $\Sh{i',i}\deh_{b,i} = \deh_{b,i'}$. 
\subsection{Quasi-particle treatment}
The quasi-particle transformation
\begin{align}
  \deh_{b,i} &= \sum_j\dc\left(u_{b,i,j}\alh_{b,j} + v^*_{b,i,j}\alhd_{b,j}\right),
\label{e:EPAPSdehuv}
\end{align}
(with the operators $\alh_{b,j}$ satisfying standard bosonic commutation relations) brings $\Kh_{2,b,i}$ into the form:
\begin{align}
 \Kh_{2,b,i} = \sum_{j,k}\dc 
&\left\{
	\alhd_{b,j}\alh_{b,k}
	\left[
		u^*_{b,i,j}\Lh_{b,i}u_{b,i,k} + \frac{U_{0b}}{2}
		\left(
			z^2_i u^*_{b,i,j}v_{b,i,k} + z_i^{*2} v^*_{b,i,j} u_{b,i,k}
		\right)
	\right]\right.\notag\\
	&+\alh_{b,j}\alhd_{b,k}
	\left[
		v_{b,i,j}\Lh_{b,i} v^*_{b,i,k} + \frac{U_{0b}}{2}
		\left(
			z_i^2 v_{b,i,j}u^*_{b,i,k} + z_i^{*2}u_{b,i,j}v^*_{b,i,k}
		\right)
	\right]\notag\\
	&+\alh_{b,j}\alh_{b,k}
	\left[
		v_{b,i,j}\Lh_{b,i}u_{b,i,k} + \frac{U_{0b}}{2}
		\left(
			z^2_i  v_{b,i,j}v_{b,i,k}  + z_i^{*2} u_{b,i,j} u_{b,i,k}
		\right)
	\right]\notag\\
	&+\left.\alhd_{b,j}\alhd_{b,k}
	\left[
		u^*_{b,i,j}\Lh_{b,i}v^*_{b,i,k} + \frac{U_{0b}}{2}
		\left(
			z^2_i u^*_{b,i,j} u^*_{b,i,k}   + z_i^{*2} v^*_{b,i,j}v^*_{b,i,k}
		\right)
	\right]\right\}.\label{e:EPAPSK2bi}
\end{align}
To calculate the tunneling term, we first consider a property of the shift operator, $\Sh{i',i}$. Since $J_{b,i,i'} = J_{b,i',i}^*$, we have:
\begin{align}
\sum_{i,i'} x_i^*J_{b,i,i'}\Sh{i',i}y_i 
&= \sum_{i,i'} \left(J_{b,i,i'}\Sh{i',i}x_i\right)^*y_{i}  ,
\end{align}
by interchanging the roles of the dummy variables\afn[.]{This result continues to apply if we exclude, e.g. beyond nearest or beyond next-nearest neighbors by symmetrically setting $J_{b,i,i'}=0$ for hopping terms not required.} From \eqref{e:EPAPSLdef}, since the diagonal terms in
$\Lh$ are real, we therefore have\afn{This result shows that $\Lh$ is Hermitian under the inner product $\av{x|y} = \sum_{i}x_i^*y_{i}$} $\sum_i{x_i^*\Lh_{b,i}y_i} = \sum_i{\left(\Lh_{b,i}x_i\right)^*y_i}$ so that:
\begin{equation}
  \sum_i{x_i^*\Lh_{b,i}y_i} = \half\left[\sum_i{\left(\Lh_{b,i}x_i\right)^*\hspace{-1mm}y_i} +\hspace{-1mm} \sum_i{x_i^*\Lh_{b,i}y_i}\right],
\end{equation}
and:
\begin{align}
 \sum_i\Kh_{2,b,i} = \half\sum_{i,j,k}\dc 
&\left[
(E_{b,j}+E_{b,k})\left( \alhd_{b,j}\alh_{b,k}u^*_{b,i,j} u_{b,i,k} 
- \alh_{b,j}\alhd_{b,k} v_{b,i,j} v^*_{b,i,k} \right)\bsplit
	-(E_{b,j}-E_{b,k})\left(\alh_{b,j}\alh_{b,k} v_{b,i,j}u_{b,i,k}-\alhd_{b,j}\alhd_{b,k} u^*_{b,i,j}v^*_{b,i,k}\right)
	\right].	\label{e:EPAPSsumkbi}
\end{align}
We choose the modes to satisfy the Bogoliubov-de Gennes equations:
\begin{align}
  \Lh_{b,i}u_{b,i,j} + U_{0b}z_i^2 v_{b,i,j} &= E_{b,j} u_{b,i,j},\label{e:EPAPSLhu}\\
  \Lh_{b,i}v_{b,i,j} + U_{0b}z^{*2}_i u_{b,i,j} &= -E_{b,j} v_{b,i,j}.\label{e:EPAPSLhv}
\end{align}
The second term in \eqref{e:EPAPSsumkbi} is directly zero for $j=k$ and from $v_{b,i,k} \times \eqref{e:EPAPSLhu} + u_{b,i,k}\times \eqref{e:EPAPSLhv}$ and applying $\Lh$ to the left:
\begin{align}
 ( E_{b,j} + E_{b,k}) (u_{b,i,j}v_{b,i,k} - v_{b,i,j}u_{b,i,k})&= 0 ,
\end{align}
so, for $j\ne k$ we have:
%
  $v_{b,i,j}u_{b,i,k} = u_{b,i,j}v_{b,i,k}$, 
%
taking $E_{b,k}$ to be non-negative \cite{Fetter72}.  Therefore, the sum of each pair of opposite off-diagonal elements of the coefficients of $\alh_{b,j}\alh_{b,k}$ is zero. The same argument works for the off-diagonal coefficients of $\alhd_{b,j}\alhd_{b,k}$ using the complex conjugate.  

The first term of \eqref{e:EPAPSsumkbi} becomes:
\begin{align}
(E_{b,j}\hspace{-1mm}+\hspace{-1mm}E_{b,k})\left[ \alhd_{b,j}\alh_{b,k}\left(u^*_{b,i,j} u_{b,i,k}\hspace{-1mm}-\hspace{-1mm} v^*_{b,i,j}v_{b,i,k}\right)\hspace{-1mm}-\hspace{-1mm} \delta_{jk} \norm{v_{b,i,j}}^2\right],
\end{align}
where we have exchanged the dummy variables $j$ and $k$ for the $\alh_{b,j}\alhd_{b,k}$ terms. 
From $u^*_{b,i,k}\times \eqref{e:EPAPSLhu} + v^*_{b,i,k}\times \eqref{e:EPAPSLhv}$ and applying $\Lh$ to the left:
\begin{align}
  (E_{b,j}-E_{b,k} ) (u_{b,i,j}u_{b,i,k}^* -v_{b,i,j}v_{b,i,k}^* ) &= 0,
\end{align}
so taking the complex conjugate for $j\ne k$ we have $u_{b,i,j}^*u_{b,i,k} =v_{b,i,j}^*v_{b,i,k}$, eliminating the off-diagonal terms, and using
$\sum_{i}\dc \left(\norm{u_{b,i,j}}^2 - \norm{v_{b,i,j}}^2\right) = 1$ for the diagonal terms, the Hamiltonian is reduced to the diagonal form:
\begin{align}
  \Kq &= 
 \sum_{i} z_i^*\left(-\sum_{i'} J_{0,i,i'} \Sh{i',i}  + \vi - \mu +  \frac{U_{00}}{2}\norm{z_i}^2\right)z_i
  + \sum_{b,j}\dc  E_{b,j}\left(\alhd_{b,j}\alh_{b,j} -  \sum_i \norm{v_{b,i,j}}^2 \right)\label{e:EPAPSKhdiag}.
\end{align}
\end{widetext}

\renewcommand{\bibname}{References}

\end{document}